%% file: main.tex
\documentclass[lettersize,journal]{IEEEtran}
\usepackage{cite}
\usepackage{bbding}
\usepackage{pifont}  
\usepackage{listings}
\usepackage[table]{xcolor}
\usepackage{amsmath,amssymb,amsfonts}
\usepackage{algorithmic}
\usepackage{graphicx}
\usepackage{textcomp}
\usepackage{hyperref}
\usepackage{xcolor}
\usepackage{float}
\usepackage{tabularx}      % 自适应宽度表格
\usepackage{booktabs}     % 专业表格线
\usepackage{siunitx}      % 数值对齐
\usepackage{multirow}     % 多行合并
\usepackage{makecell}     % 单元格格式控制
\usepackage{array}
\usepackage{diagbox}
\usepackage{tabularx}
\usepackage{adjustbox}
\usepackage{makecell}
\usepackage{xcolor}
\usepackage[many]{tcolorbox}
\usepackage[numbers,sort&compress]{natbib}
\usepackage{graphicx}      % 基本的图形支持
\usepackage{balance}
\usepackage{xspace}

\usepackage{tcolorbox}
\newcommand{\rqbox}[1]{\begin{tcolorbox}[left=4pt,right=4pt,top=4pt,bottom=4pt,colback=gray!5,colframe=gray!40!black,before skip=6pt,after skip=6pt]#1\end{tcolorbox}}

\definecolor{greenbg}{RGB}{220, 255, 220}
\definecolor{redbg}{RGB}{255, 220, 220}
\definecolor{graybg}{RGB}{240, 240, 240}
\newcommand{\bench}{LibRec\xspace}

\newcommand{\data}{LibEval\xspace}

\newcolumntype{C}[1]{>{\centering\arraybackslash}p{#1}}

\newcolumntype{L}[1]{>{\raggedright\arraybackslash}m{#1}}

\def\BibTeX{{\rm B\kern-.05em{\sc i\kern-.025em b}\kern-.08em
    T\kern-.1667em\lower.7ex\hbox{E}\kern-.125emX}}

\begin{document}

\title{\bench: Benchmarking Retrieval-Augmented LLMs for Library Migration Recommendations}
%\title{Target Library Recommendations Using LLMs}

\author{Junxiao Han, Yarong Wang, Xiaodong Gu, Cuiyun Gao, Yao Wan, Song Han, David Lo, and Shuiguang Deng
\IEEEcompsocitemizethanks{\IEEEcompsocthanksitem  Junxiao Han and Song Han are with the School of Computer and Computing Science, Hangzhou City University, Hangzhou, China. E-mail: hanjx@hzcu.edu.cn and hans@hzcu.edu.cn.
\IEEEcompsocthanksitem Yarong Wang is with the Polytechnic Institute, Zhejiang University, Hangzhou, China. E-mail: wangyarong@zju.edu.cn
\IEEEcompsocthanksitem Xiaodong Gu is with the School of Computer Science, Shanghai Jiao Tong University, Shanghai, China. E-mail: xiaodong.gu@sjtu.edu.cn
\IEEEcompsocthanksitem Cuiyun Gao is with the School of Computer Science and Technology, Harbin Institute of Technology, Shenzhen, China. E-mail: gaocuiyun@hit.edu.cn
\IEEEcompsocthanksitem Yao Wan is with the College of Computer Science and Technology, Huazhong University of Science and Technology, Wuhan, China. E-mail: wanyao@hust.edu.cn
\IEEEcompsocthanksitem David Lo is with the School of Computing and Information Systems, Singapore Management University, Singapore. E-mail: davidlo@smu.edu.sg
\IEEEcompsocthanksitem Shuiguang Deng is with the College of Computer Science and Technology, Zhejiang University, Hangzhou 310027, China. E-mail: dengsg@zju.edu.cn
\IEEEcompsocthanksitem Junxiao Han and Yarong Wang contribute equally.
\IEEEcompsocthanksitem Shuiguang Deng is the corresponding author.
}}

 \pagenumbering{arabic}

\maketitle
 
%\IEEEtitleabstractindextext{
\begin{abstract}

\emph{Library migration} refers to replacing outdated or unsuitable libraries with more appropriate alternatives. Although substantial research has been conducted on the empirical analysis of library migration, an automated framework for recommending alternative libraries is still lacking. Such a framework could substantially reduce developer effort and enhance the efficiency of the library migration process. Meanwhile, large language models (LLMs) have demonstrated exceptional performance across a wide range of software engineering (SE) tasks. However, their potential for addressing the library migration recommendation task remains unexplored. In this paper, we propose \bench, a novel framework that integrates the capabilities of LLMs with retrieval-augmented generation (RAG) techniques to automate the recommendation of alternative libraries. The framework further employs in-context learning to extract migration intents from commit messages to enhance the accuracy of its recommendations. To evaluate the effectiveness of \bench, we introduce \data, a benchmark designed to assess the performance in the library migration recommendation task. \data comprises 2,888 migration records associated with 2,368 libraries extracted from 2,324 Python repositories. Each migration record captures source-target library pairs, along with their corresponding migration intents and intent types. Based on \data, we evaluated the effectiveness of ten popular LLMs within our framework, conducted an ablation study to examine the contributions of key components within our framework, explored the impact of various prompt strategies on the framework's performance, assessed its effectiveness across various intent types, and performed detailed failure case analyses. Our experimental results reveal several key findings: (1) \bench demonstrates strong effectiveness in recommending target libraries, with the Claude‑3.7‑Sonnet model achieving the best performance across all metrics; (2) both components in the framework are critical for achieving optimal performance; (3) The One-shot prompt strategy emerges as the most robust and reliable paradigm across six LLMs, consistently delivering superior performance; and (4) the framework demonstrates superior performance for intent types related to issues with source libraries. These findings provide valuable insights into the capabilities of various LLMs in our framework for target library recommendation, highlight promising directions for future research, and provide practical suggestions for researchers, developers, and users.

\end{abstract}


%\maketitle

\section{Introduction}\label{intro}
\input{intro}

\section{Related Work}\label{relatedwork}
\input{related}

\section{\bench: Retrieval-Augmented Library Recommendation}\label{LibRec}
\input{LibRec}

\section{\data: The Benchmark}\label{LibEval}
\input{LibEval}

\section{Experimental Setup}\label{setup}
\input{setup}

\section{Evaluation Results}\label{result}

\subsection{RQ1: Overall Performance}\label{rq1}
\input{rq1}

% rq1的表格
\subsection{RQ2: Ablation Study}\label{rq2}
\input{rq2}

\subsection{RQ3: Effectiveness of Different Prompt Engineering Strategies}\label{rq3}
\input{rq3}

\subsection{RQ4: Effectiveness across Different Intent Types}\label{rq4}
\input{rq4}
\subsection{RQ5: Case Study}\label{rq5}
\input{rq5}

%Wang \section{Discussion}\label{discussion}
%Wang \input{discussion}

\section{Discussion}\label{implication}
\input{implication}

\section{Threats To Validity}\label{threat}
\input{threat}

\section{Conclusion}\label{conclusion}
\input{conclusion}

%\section*{Acknowledgment}

\balance
%\normalem
% \bibliographystyle{IEEEtran}
% \bibliography{reference}
\input{references}

\end{document}

%% file: intro.tex
% LLM的发展动态，并未应用于library migration推荐，依赖于第三方库 第三方库重要性 由于各种原因需要替换 引出 library migration  手动过程是耗时的，所以自动实现非常需要。
% 目前的related work大部分target在Java，或者target在API mapping。仅有的针对library recommendation的work也仅是使用metric进行target library的筛选，数据集是固定的且数据量较小，对于未出现过的library无法得到有效推荐。少量引用LLM进行library migration的work，其中A集中在LLM生成的migration code的empirical分析，而B集中在LLM生成的code中推荐的library的empirical analysis，并未有work利用LLM的能力进行library recommendation。那么LLM在library推荐这个task上的能力这些均为解决

\begin{figure*}[!ht]
  \centering
\includegraphics[width=\textwidth,keepaspectratio]{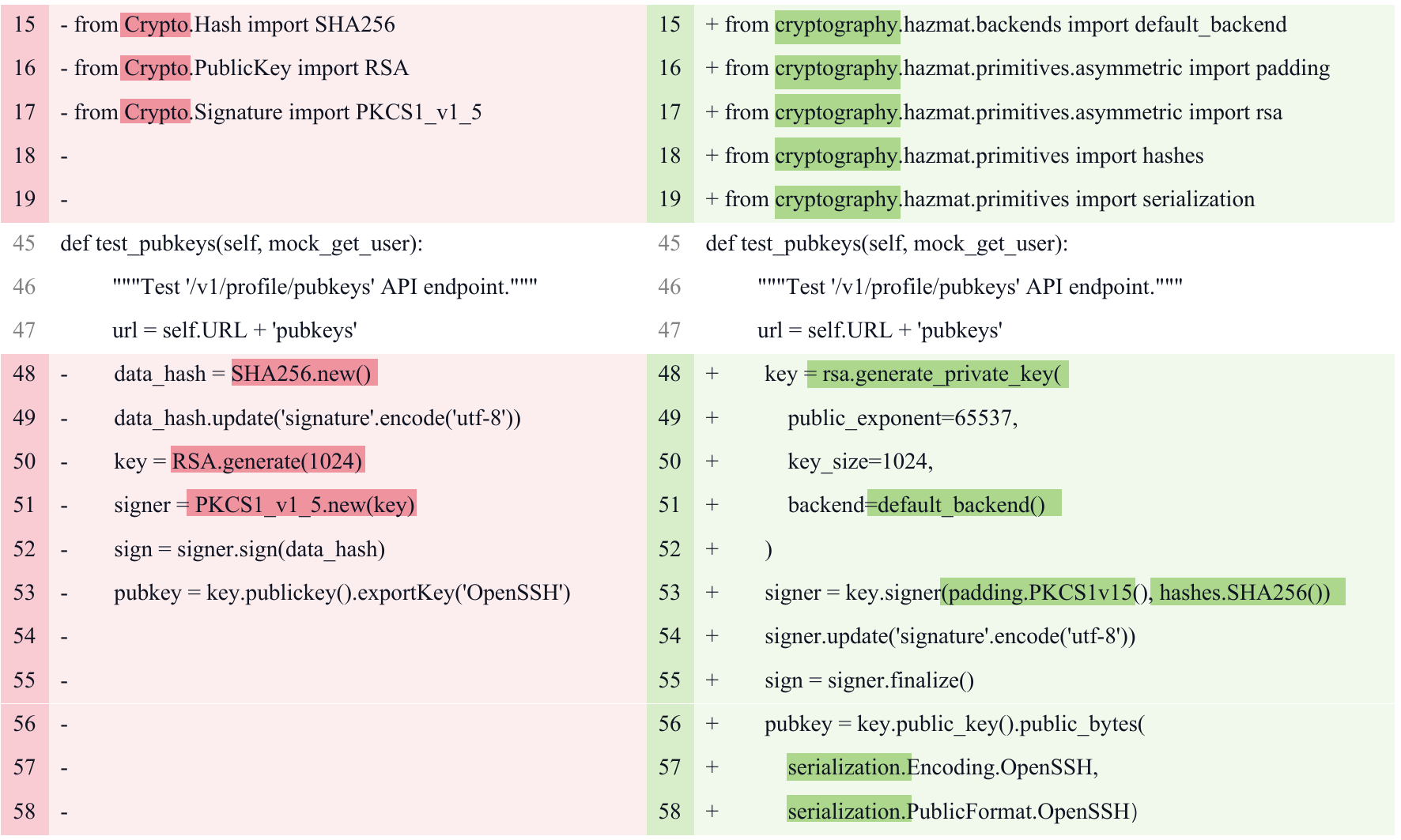}
  \caption{Library migration from \texttt{pycryptodome/Crypto} to \texttt{cryptography} extracted from \emph{svtplay-dl}, where the pink-highlighted parts represent the source library and its corresponding code to be migrated, while the green-highlighted parts denote the target library and its replacement code.}
  \label{intro}
\end{figure*}

%Large Language Models (LLMs) are increasingly being employed to support various software engineering (SE) tasks, such as code generation \cite{li2025ddpt,jiang2024survey}, code completion \cite{eghbali2024hallucinator,liu2024graphcoder}, test cases generation \cite{deljouyi2024leveraging,chen2024chatunitest}, vulnerability detection \cite{yildiz2025benchmarking,lin2025large}, and API recommendation \cite{chen2024apigen}. However, to the best of our knowledge, their application to the task of library migration has not yet been explored. 
\IEEEPARstart{L}{ibrary} migration is an essential activity in modern software development, as modern software heavily relies on third-party libraries \cite{wang2020empirical}. However, these libraries may become obsolete over time \cite{he2021multi,wang2020exploring}, introduce vulnerabilities or bugs that negatively impact dependent applications \cite{kula2018developers,ossendrijver2022towards,islam2024characterizing}, or be replaced by newer, more efficient, or easier-to-use alternatives \cite{larios2020selecting,he2021multi}. Consequently, developers often face the need to replace an outdated or unsuitable library (i.e., source library) with a more appropriate alternative (i.e., target library), a process commonly referred to as \emph{library migration}.

Despite its significance, library migration remains a complex, time-consuming, and error-prone task. Developers often dread this process, particularly in large codebases where the source library is deeply integrated \cite{kula2018developers}. Identifying appropriate alternatives automatically holds the potential to significantly reduce developer effort and improve migration efficiency. However, most existing research on library migration has been empirical, focusing on the frequency, domains, mechanisms, and rationales behind library migration \cite{teyton2014study,he2021large,gu2023self,islam2024characterizing}. A limited body of work has investigated library migration recommendations, primarily focusing on applying several metrics to screen existing migration rules for generating recommendations for source libraries \cite{he2021multi,zhang2024multi,mujahid2023go}. Unfortunately, these approaches lack the capability to automatically recommend suitable target libraries, leaving developers to manually identify alternatives. Moreover, they are unable to recommend libraries that are not covered by existing migration rules. 

In recent years, Large Language Models (LLMs) are increasingly being employed to support various software engineering (SE) tasks, such as code generation \cite{li2025ddpt,jiang2024survey}, code completion \cite{eghbali2024hallucinator,liu2024graphcoder}, test cases generation \cite{deljouyi2024leveraging,chen2024chatunitest}, vulnerability detection \cite{yildiz2025benchmarking,lin2025large}, and API recommendation \cite{chen2024apigen}. This suggests that LLMs may also possess the capability to perform library migration recommendations. However, to the best of our knowledge, their application to library migration recommendation has not yet been explored. Existing LLM-based techniques for library migration tasks have primarily focused on API recommendations rather than library migration. For instance, ToolLLM facilitates the recommendation of chains of API calls for given instructions \cite{qin2023toolllm}, Gorilla employs a fine-tuned LLaMA model for API recommendation \cite{patil2024gorilla}, and APIGen leverages generative models with enhanced in-context learning for API suggestions \cite{chen2024apigen}. Although LLMs demonstrate the potential in API recommendation tasks, their application to library migration remains unexplored. 

To bridge this gap, we propose \bench, a framework that combines the capabilities of LLMs with retrieval-augmented generation (RAG) techniques \cite{lewis2020retrieval,guu2020retrieval} to automate library migration recommendations. The framework leverages in-context learning of migration intents extracted from commit messages to enhance library recommendations. For instance, consider the migration of the Python library \texttt{pycryptodome/Crypto} to its analogous library \texttt{cryptography}, as illustrated in Figure \ref{intro}. In this case, the developer stated in their commit message, ``This will make it easier to support older distros. For example, Ubuntu 16.04 and Debian stable (9)." In such scenarios, our framework is designed to recommend the target library, \texttt{cryptography}, based on developer-provided instructions, the names of source libraries (e.g., \texttt{pycryptodome/Crypto}), and migration intents expressed in commit messages.

To assess the effectiveness of \bench, we construct \data, a benchmark specifically designed to evaluate its performance in target library recommendation tasks. \data comprises 2,888 migration records associated with 2,368 libraries extracted from 2,324 Python repositories, each record includes the source-target library pairs, along with their generated migration intents and intent types. Building on this foundation, we evaluated the effectiveness of ten popular LLMs within our framework, including general LLMs of GPT-4o-mini, GPT-4, DeepSeek-V3, Qwen-Plus, Llama-3.3, and Claude-3.7-Sonnet; code LLMs such as Qwen-Coder-Plus and Qwen2.5-Coder from the Qwen family; and reasoning LLMs of DeepSeek-R1 and QwQ-Plus (RQ1). Additionally, we conducted an ablation study to examine the contributions of key components within the framework, specifically the retrieval-augmented component and the inclusion of migration intents in prompts (RQ2). To deepen our understanding, we explored various prompt strategies, including One-shot and Chain-of-Thought (CoT), to assess their impact on the framework's performance (RQ3). Additionally, we assessed the framework's effectiveness across various intent types, such as issues with source libraries, advantages of target libraries, and project-related reasons (RQ4), and performed case studies to analyze failure cases (RQ5).   

Based on our experimental results, we summarize the following main findings: 1) \bench demonstrates strong effectiveness in recommending target libraries, with the Claude‑3.7‑Sonnet model achieving the best performance across all metrics. 2) Both the retrieval-augmented component and migration intents within the framework are essential for achieving optimal effectiveness. On average, retrieval augmentation exhibits greater significance compared to migration intents. 3) The one-shot prompt strategy emerges as the most robust and reliable paradigm across six models, consistently delivering superior performance. In contrast, the COT prompt strategy excels primarily in reasoning LLMs and high-capacity models like Claude-3.7-Sonnet. 4) Our framework demonstrates superior performance for intent types related to issues with source libraries. 5) Most recommendation failures occur within the intent type of advantages of target libraries, with the sub-category \emph{Enhanced Features} (i.e., more specific features) accounting for the largest proportion of failures in this group.

In summary, this paper makes the following contributions:

\begin{itemize}
    \item We propose \bench, a framework that combines the capabilities of LLMs with RAG techniques, and leverages in-context learning of migration intents to recommend target libraries automatically.
    \item We introduce \data, a benchmark designed for the target library recommendation task, comprising 2,888 records of source-target library pairs, along with their corresponding generated migration intents and intent types.
    \item We evaluate the performance of ten popular LLMs in our framework, providing insights into their effectiveness in the target library recommendation task, and reveal the contributions of each key component in our framework. Moreover, we examine the performance of our framework across various intent types, offering a detailed understanding of its effectiveness in diverse migration scenarios.
\end{itemize}

The remainder of this paper is organized as follows. Section \ref{relatedwork} reviews the related work. Section \ref{LibRec} proposes the framework for library migration recommendations, while Section \ref{LibEval} introduces the benchmark for evaluation. Section \ref{setup} presents the experimental setup, and Section \ref{result} reports the findings corresponding to each of the five research questions. In Section \ref{implication}, we discuss the implications of our findings to researchers, developers, and users. Section \ref{threat} addresses potential threats to the validity of our study, and Section \ref{conclusion} concludes this paper and outlines directions for future work.

%% file: related.tex
\subsection{Traditional Library Migration}

Several studies have conducted empirical analyses of library migrations or proposed traditional approaches to recommend libraries to developers. Teyton et al. \cite{teyton2014study} analyzed Java open-source software to investigate the frequency, contexts, and reasons for library migration. He et al. \cite{he2021large} examined 19,652 Java GitHub projects to understand how and why library migration happens, and derived 14 frequently mentioned
migration reasons (e.g., lack of maintenance, usability, and integration) via thematic analysis of related commit messages, issues, and pull requests. Gu et al. \cite{gu2023self} compared the frequency, domain, rationales, and unidirectionality of self-admitted library migrations (SALMs) across Java, JavaScript, and Python ecosystems, revealing domain similarity across these ecosystems, and revealing differences in rationale distributions, ecosystem-specific domains, and levels of unidirectionality. Islam et al. \cite{islam2024characterizing} conducted an empirical analysis of Python library migrations, identifying a taxonomy of PyMigTax to classify migration-related code changes, which consisted of 3,096 migration-related code changes across 335 Python library migrations from 311 client repositories spanning 141 library pairs in 35 domains. Moreover, He et al. \cite{he2021multi} proposed an approach for recommending library migrations in Java projects by leveraging multiple metrics, including Rule Support, Message Support, Distance Support, and API Support. Based on the metrics in this work, Zhang et al. \cite{zhang2024multi} incorporated the metric of label correlations of libraries in the Maven Central Repository to enhance migration recommendations. Mujahid et al. \cite{mujahid2023go} developed an approach to identify declining packages and suggest alternative libraries based on observed migration patterns, resulting in 152 dependency migration suggestions. 

Most existing research is predominantly conducted on Java datasets, while the exploration of Python datasets is limited. Moreover, current studies primarily focus on analyzing the frequency and distribution of reasons for library migrations, without providing concrete recommendations for target libraries to replace source libraries. The few studies addressing library migration recommendations typically rely on multiple metrics to filter existing migration rules, but fall short of incorporating the latest knowledge. To address these gaps, our study establishes a benchmark for the library migration task and proposes a target library recommendation framework that leverages the capabilities of LLMs to deliver effective target library recommendations.

\subsection{LLM-based library migration}

The majority of LLM-based techniques for library migration primarily focus on recommending APIs or API sequences. Qin et al. \cite{qin2023toolllm} proposed ToolLLM, a framework encompassing data construction, model training, and evaluation, designed to leverage ChatGPT for identifying valid solution paths (i.e., chains of API calls) for given instructions. Nan et al. \cite{nan2024ddasr} introduced DDASR, a framework for recommending API sequences that integrate both popular and tail APIs. Their framework leveraged LLMs for learning query representations, and clustered tail APIs with similar functionality, replacing them with cluster centers to enhance the understanding of tail APIs. Experimental results demonstrate that DDASR achieves superior diversity in recommendations while maintaining high accuracy. 

Moreover, Wang et al. \cite{wang2024prompt} proposed PTAPI, a novel approach designed to enhance API recommendation performance by visualizing users' intentions based on their queries. The method began by identifying a prompt template from Stack Overflow posts based on the user's input, which was then combined with the input to generate a refined query. This refined query leveraged dual information sources from Stack Overflow posts and official API documentation to provide more precise recommendations. Yang et al. \cite{yang2025dskipp} proposed DSKIPP, a novel prompt method designed to enhance LLMs for Java API recommendations. This approach systematically guides LLMs through a hierarchical process encompassing package, class, and method levels, and incorporates development scenarios, key knowledge, and a self-check mechanism to ensure reliable results. Patil et al. \cite{patil2024gorilla} introduced Gorilla, an API recommendation tool built on a fine‑tuned LLaMA model that works with a document retriever, achieving greater flexibility and generalization compared to traditional embedding‑based techniques. Chen et al. \cite{chen2024apigen} developed APIGen, a generative API recommendation approach that employs enhanced in-context learning to suggest APIs effectively with a few examples, utilizing diverse example selection and guided reasoning to improve recommendation accuracy and interoperability. Additionally, Latendresse et al. \cite{latendresse2024chatgpt} utilized ChatGPT to generate Python code for a substantial set of real-world coding problems derived from Stack Overflow, analyzing the characteristics of libraries employed in the generated code.

%% file: LibRec.tex
\begin{figure*}[!tbp]
  \centering
\includegraphics[width=\textwidth,keepaspectratio]{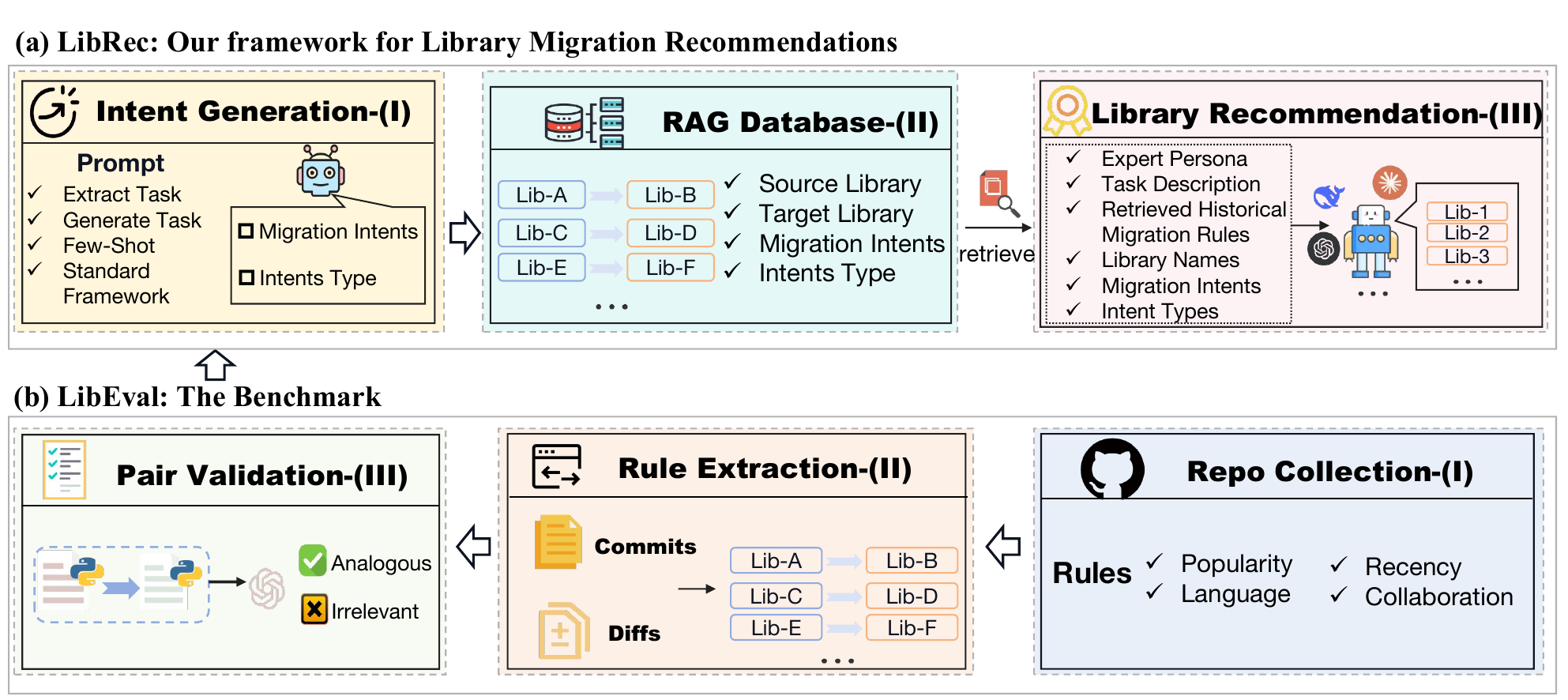}
  \caption{Overview of study workflow.}
  \label{workflow}
\end{figure*}

Figure~\ref{workflow} (a) presents the workflow of our proposed framework, \bench, which combines the capabilities of LLMs with the RAG strategy \cite{lewis2020retrieval,guu2020retrieval} to automate library migration recommendations. The framework begins by generating migration intents for each library migration pair, thereby providing additional contextual information to facilitate library migration recommendation tasks (Figure~\ref{workflow} (a)-I). Next, we construct an RAG database to support the formulation of the target library recommendation task (Figure~\ref{workflow} (a)-II). Finally, LLMs, enhanced with RAG techniques, and guided by migration intents embedded in prompts, enable the recommendation of appropriate target libraries (Figure~\ref{workflow} (a)-III). A detailed explanation of this framework is provided below.

\subsection{Migration Intents Generation}\label{intent}

\begin{figure}[h]
  \centering
  \includegraphics[width=1.0\columnwidth]{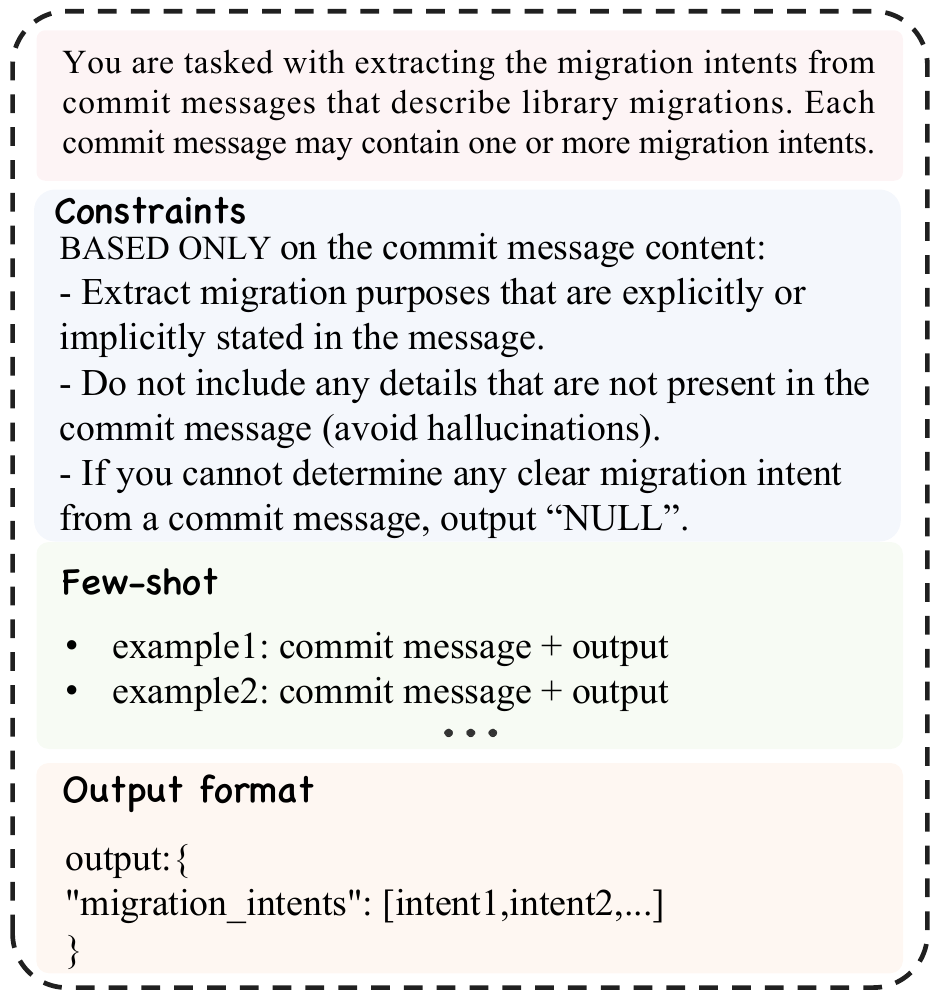}
  \caption{The prompt template used for generating migration intents.}
  \label{generateprompt}
\end{figure}

Previous studies have adopted commit messages as proxies for programmer intents to augment the capabilities of LLMs in automatic bug fixing \cite{zhang2025patch}. Inspired by this work, we incorporate migration intents to enhance the performance of LLMs in library recommendation. However, a substantial amount of commit messages from open-source projects lack critical information or are even empty \cite{eliseeva2023commit}. Additionally, many commit messages cover a wide range of subjects \cite{choshen2021comsum}. In this regard, just adopting commit messages as proxies for migration intents can introduce noise. To address this issue and ensure the provision of high-quality migration intents, we leverage the capabilities of LLMs to extract refined migration intents from commit messages.

Specifically, we utilize an LLM (GPT-4o) to generate migration intents based on commit messages. We design a prompt template (Figure \ref{generateprompt}) employing a few-shot strategy to instruct the model to generate migration intents under specific constraints: 1) extract migration purposes explicitly or implicitly stated in the message, 2) avoid including any details not present in the commit message to mitigate the risk of hallucinated outputs, and 3) if no clear migration intent can be determined from the commit message, output ``NULL".

%Wang we use few-shot prompting with GPT-4o to mine migration intents from these messages. Specifically, the model is presented with a small set of carefully chosen examples and then asked to identify and isolate the intent expressed in each commit message. By sourcing all intent information exclusively from the original messages, we substantially mitigate the risk of hallucinated outputs. Since high-quality messages are scarce, we ultimately extract clear intents from 2,888 commit messages.

\subsection{Intent Types Determination}\label{intent-type} 

Subsequently, we classified the migration intents into distinct categories following the classification framework proposed by He et al. \cite{he2021large}. This framework defines four primary categories of \emph{Source Library Issues}, \emph{Target Library Advantages}, \emph{Project Specific Reasons}, and \emph{Other}, along with 14 associated subcategories. Intent types under \emph{Source Library Issues} typically arise due to issues with the source library, such as the lack of maintenance or being outdated (i.e., subcategory of \emph{Not Maintained/Outdated}), security vulnerabilities (i.e., \emph{Security Vulnerability}), or bugs and defect issues (i.e., \emph{Bug/Defect Issues}). Intents classified under \emph{Target Advantages} reflect the benefits provided by the target library, including better usability (i.e., \emph{Usability}), more specific features (i.e., \emph{Enhanced Features}), superior functional performance (i.e., \emph{Performance}), smaller size/complexity (i.e., \emph{Size/Complexity}), higher popularity (i.e., \emph{Popularity}), stronger stability and maturity (i.e., \emph{Stability/Maturity}), and more community activities (i.e., \emph{Activity}). For intents categorized as \emph{Project Specific}, the reasons are tied to the unique needs of the project, such as easier integration (i.e., \emph{Integration}), project simplification (i.e., \emph{Simplification}), ensuring license compatibility (i.e., \emph{License}), and organization influences (i.e., \emph{Organization Influence}). 

\begin{figure}[!t]
  \centering
  \includegraphics[width=\linewidth]{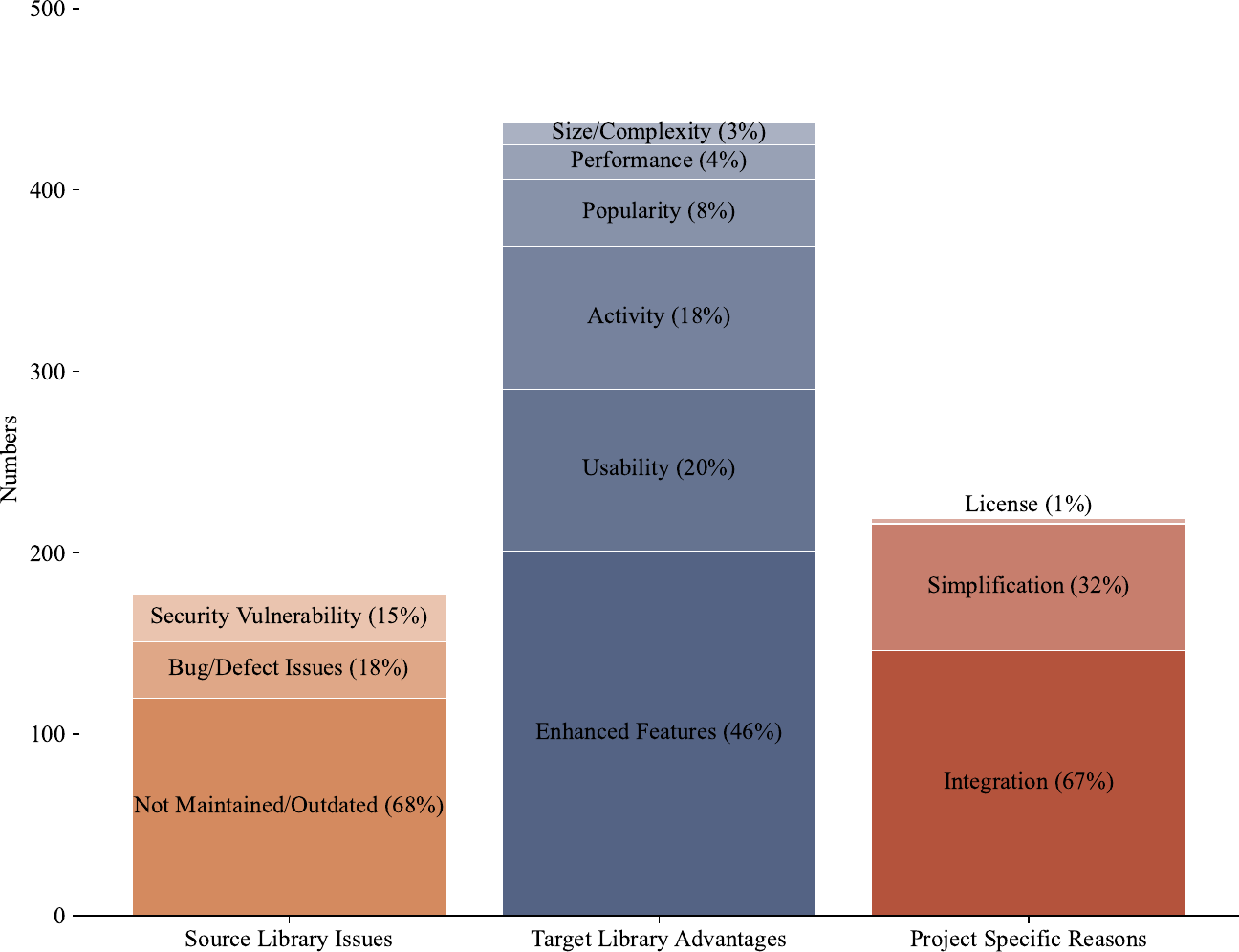}
  \caption{The distribution of migration intents.}
  \label{wenshifenlei}
\end{figure}

We employ an LLM (GPT-4o) with few-shot prompting to automatically classify each migration intent into its corresponding intent types, and Figure \ref{wenshifenlei} presents the distribution of the three primary categories (excluding the ``Other'' type) of migration intent types. Within the \emph{Source Library Issues} category, the subcategory \emph{Not Maintained/Outdated} accounts for more than half of the intent types. Similarly, \emph{Integration} constitutes the majority of intent types within the \emph{Project Specific} category, while \emph{Enhanced Features} represents nearly half of the intent types under \emph{Target Advantages}.   %and the overlaps between pairs of categories indicate that, in certain cases, migrations are driven by multiple intents. 

\subsection{Construction of RAG Database}\label{RAG}

As depicted in Figure \ref{workflow}, our goal is to recommend suitable target libraries for a given source library. To achieve this, we first construct an RAG database, where each entity is represented as $H=\{(S_j,T_j,I_j,C_j)\}_{j=1}^{|H|}$, forming a 4-tuple library migration dataset. Here, $S_j$ denotes the $j\text{-}th$ source library, $T_j$ represents its corresponding target library, $I_j$ indicates the set of migration intents, and $C$ specifies the intent types. The recommendation process begins by retrieving migration entities similar to the given source library from the constructed RAG database. For this purpose, we employ the sparse keyword-based BM25 score \cite{robertson2009probabilistic} as our retrieval metric. BM25, a probabilistic model widely adopted in prior research \cite{wei2020retrieve,li2021editsum,zhang2025patch}, operates as a bag-of-words retriever that estimates lexical-level similarity between two sentences. Higher BM25 scores indicate greater similarity between the sentences. 

Specifically, each query is formulated as a triple $q = (s,i,c)$, where $s$ denotes the given source library, $i$ represents the associated migration intents, and $c$ indicates the intent types. Based on this multi-faceted library migration context, \bench retrieves the top-$k$ most relevant migration entities from the RAG database for a given query $q$. 

%Wang In the library migration task, the retrieval stage aims to guide migration by finding similar patterns in historical migration rules. To achieve this goal, we employ the sparse keyword–based BM25 score\cite{robertson2009probabilistic} as the retrieval metric. BM25 is a probabilistic model that scores document-query pairs using term frequency and inverse document frequency. It treats text as a bag of words and has been applied in code retrieval and generation tasks\cite{wei2020retrieve,li2021editsum}. A higher BM25 score indicates greater lexical similarity between two texts. Let $H=\{(S_j,T_j,I_j,C_j)\}_{j=1}^{|H|}$ represent a library migration datasets containing $|H|$ 4-tuple library migration rules. Where $S_j$ denotes the $j\text{-}th$ source library, $T_j$ denotes its target library, $I_j$ denotes the set of migration intents, and $C$ is the intent categories. We represent each retrieval query as the triple $q = (S,I,C)$ and the retriever returns the top-$k$ historical rules most similar to $q$. The LLM is then able to learn migration patterns from the retrieved similar migration rules.

\subsection{Target Library Recommendation}\label{TLR}

\begin{figure}[!t]
  \centering
  \includegraphics[width=\linewidth]{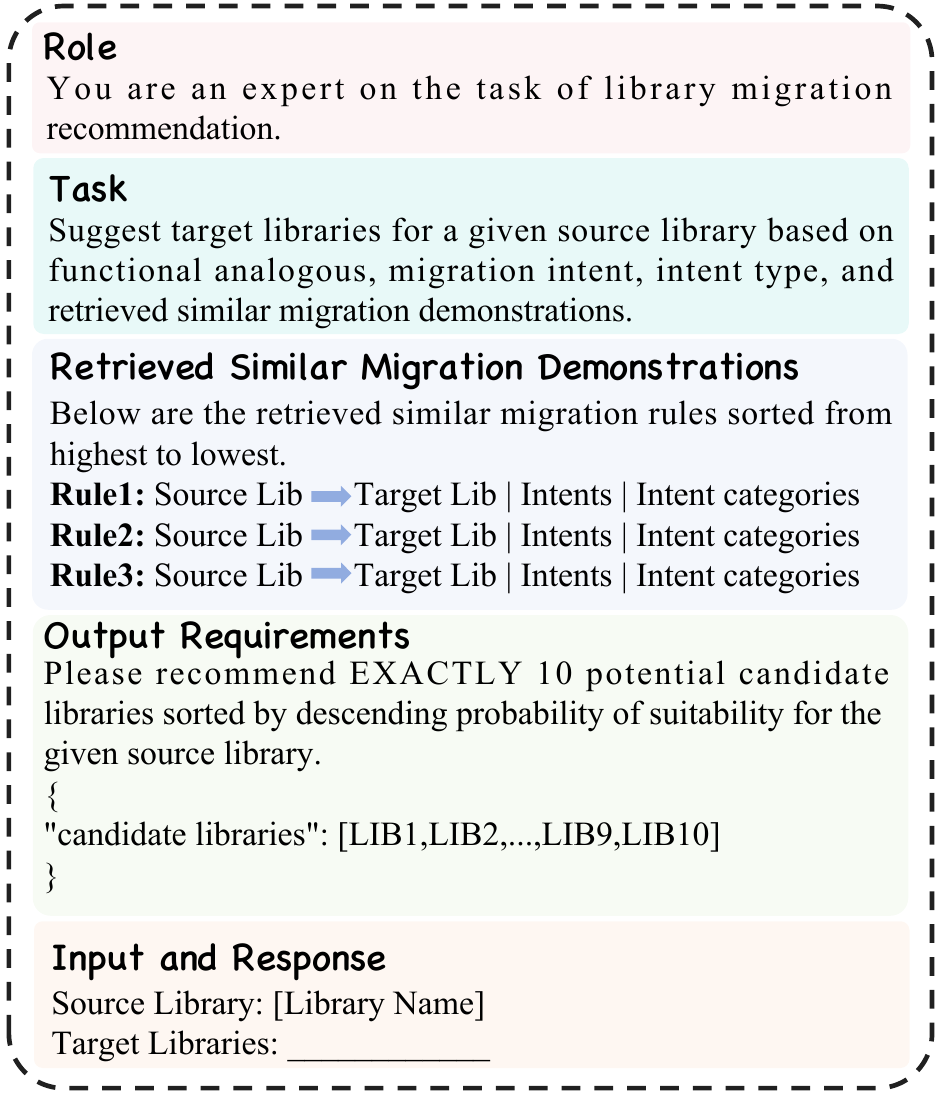}
  \caption{The prompt template used for target library recommendation.}
  \label{promptRec}
\end{figure}

\bench leverages the capabilities of LLMs and employs a structured prompt (as shown in Figure \ref{promptRec}) to require LLMs to recommend appropriate target libraries. Specifically, \bench incorporates similar migration demonstrations retrieved from the constructed RAG database (as detailed in Section \ref{RAG}). These demonstrations provide LLMs with relevant examples of similar library migrations. Additionally, migration intents and intent types are incorporated to augment the capabilities of LLMs in precise target library recommendations. In particular, the structured prompt includes the name of the source library, its similar migration demonstrations, its corresponding migration intents, and intent types, providing LLMs a comprehensive understanding of the library migration context. \bench instructs LLMs to output a ranked list of the top ten candidate libraries, ordered by their estimated suitability for the given source library. This scenario enhances development efficiency and reduces the time developers spend searching and selecting appropriate target libraries.

%Wang In this phase, the LLM is provided with a single, structured prompt. The prompt inputs consist of: (1) a role specification positioning the model as an expert software migration consultant; (2) a task description directing it to identify and rank suitable target libraries; (3) the retrieved migration rules, sorted from highest to lowest similarity; (4) metadata for the source library; (5) the migration intents; and (6) the migration intent categories. 

%Wang \paragraph{Target Library Masking}

%Wang During the analysis of the test dataset, we identified potential data leakage risks in migration intents extracted from commit messages. These intents occasionally contained explicit target library names or variants, which might inadvertently reveal ground truth information when prompting LLMs. To address this, we manually masked all target library references by replacing them with [MASK] tokens. This ensures the LLM generates recommendations without accessing privileged target details, thereby preserving evaluation integrity.

%Wang \paragraph{Recommendation List Generation}

%% file: LibEval.tex
To evaluate the effectiveness of \bench, we construct \data, a benchmark specifically designed to assess its performance in target library recommendation tasks. Specifically, we outline the construction process of \data (as illustrated in Figure~\ref{workflow} (b)), covering data collection (Section \ref{repo} to \ref{analogous}) and data generation (Section \ref{generation} and \ref{mask}).

\subsection{Repository Collection}\label{repo}

Since Python became the most popular programming language on GitHub after 2024, driven by its widespread adoption in AI-related domains \cite{staff2024octoverse}, our study focuses specifically on Python repositories. According to the findings of Coelho et al. \cite{coelho2020github}, 50\% of GitHub projects become unmaintained after 50 months (approximately 4 years). To ensure the timeliness of our dataset, we filtered and retained GitHub repositories with at least one commit made after January 1, 2020. To safeguard the quality of our dataset, we performed a filtering process. Drawing inspiration from prior studies \cite{he2021large,han2024sustainability}, we excluded forked repositories, as well as repositories with fewer than 10 stars or fewer than 2 contributors. Ultimately, we collected 62,986 Python repositories. 
%up to March 2025.

\subsection{Candidate Migration Commit Identification}

We analyze all commits in the collected repositories to identify candidate migration commits. To achieve this, we adopt heuristics established in prior studies \cite{islam2023pymigbench}, enabling us to exclude non-migration commits effectively. The heuristics are: 1) \textbf{Exclude merge commits}, we exclude merge commits to avoid redundant analysis of changes already present in parent branches. 2) \textbf{Dependencies have changed}, since migration involves the replacement of one library with another, we focus on commits that include both additions and deletions of lines in dependency files, as this serves as clear evidence of a library migration. Here, for instance, we consider \emph{requirement.txt} as the dependency file. 3) \textbf{Exclude bot-created commits}, we exclude them as they typically relate to version updates and do not involve code modifications.

\begin{comment}
\begin{itemize}
    \item Exclude merge commits: We systematically filter out merge commits to eliminate redundant analysis of changes that already exist in parent branches. This heuristic ensures each code modification is only processed once during our migration detection pipeline. 
    \item Must have dual dependency changes: We only keep commits that have both additions and deletions of lines in the dependency file (e.g., requirements.txt), as this clearly proves a library migration occurred.
    \item Exclude bot-created commits: We exclude commits generated by bots, as these typically involve dependency version updates rather than developer-initiated library migrations.
\end{itemize}
\end{comment}

For the first two heuristics, we utilize PyDriller\footnote{\url{https://pydriller.readthedocs.io}} to perform necessary checks. For the third heuristic, we identify commit authors whose names contain the term ``bot" to exclude bot-created commits. Following this process, we identified 433,602 candidate migration commits, along with their commit messages, from 44,288 repositories. %The remaining repositories contained no candidate migration commits.

\subsection{Candidate Migrations Extraction}

\begin{lstlisting}[language=Python, caption={Sample library replacements in a requirements.txt file}, label=list1]
(*@\colorbox{greenbg}{\strut\parbox{\dimexpr\linewidth-2\fboxsep}{\texttt{+cryptography>=1.0,!=1.3.0}}}@*)
 gunicorn==18
 oslo.config>=1.6.0 # Apache-2.0
 oslo.db>=1.4.1 # Apache-2.0
 oslo.log
 pecan>=0.8.2
 pyOpenSSL>=0.14
(*@\colorbox{redbg}{\strut\parbox{\dimexpr\linewidth-2\fboxsep}{\texttt{-pycrypto>=2.6}}}@*)
 requests>=2.2.0,!=2.4.0
 requests-cache>=0.4.9
 jsonschema>=2.0.0,<3.0.0
\end{lstlisting}

We first automatically analyze the patch diffs of each of the 433,602 candidate migration commits to identify potential library replacements, i.e., candidate migrations. Listing \ref{list1} illustrates an example of library replacements in a dependency file, where 1 library \emph{cryptography} is added, and 1 library \emph{pycrypto} is removed. To determine the sets of removed libraries and added libraries, we adopted the approach mentioned in previous studies \cite{mujahid2023go}. Specifically, we only retained library replacements where the difference between the number of added libraries ($L_a$) and removed libraries ($L_r$) satisfies the condition that the absolute difference $\lvert L_a - L_r\rvert$ is minimal (i.e., $\lvert L_a - L_r\rvert \le 1$). Additionally, we exclude library replacements involving a large number of added and removed dependencies, defined as $L_a + L_r > m$, where $m$ is the median value of $L_a + L_r$ across all commits. This filtering was applied because a large number of library replacements typically indicates significant code refactoring rather than a simple dependency replacement. It is important to note that one commit may involve multiple pairs of library replacements, potentially resulting in multiple migrations within the same commit. However, since the median value of $m$ for $L_a + L_r$ was calculated to be 2, in this regard, we retained only those commits that contained a single pair of added-removed (i.e., source-target) libraries. Furthermore, we ignore library versions in this paper to focus exclusively on migrations between different libraries. As a result, we identified a total of 14,608 candidate migration rules.

\subsection{Non-Analogous Library Pairs Screening}\label{analogous}

In the real world, developers often replace a library with another that provides the same or similar functionalities \cite{chen2019s,gu2023self}. Accordingly, for a migration to be considered valid, it must involve an analogous pair of libraries, i.e., the added and removed libraries should offer comparable functionalities. However, manually reviewing and classifying library pairs is time-consuming and labor-intensive. Motivated by the advanced capabilities of LLMs in natural language processing \cite{geng2024large} tasks, we leverage an LLM (GPT-4o) to automatically verify whether a given library pair is analogous.

\begin{figure}[h]
  \centering
  \includegraphics[width=1.0\columnwidth]{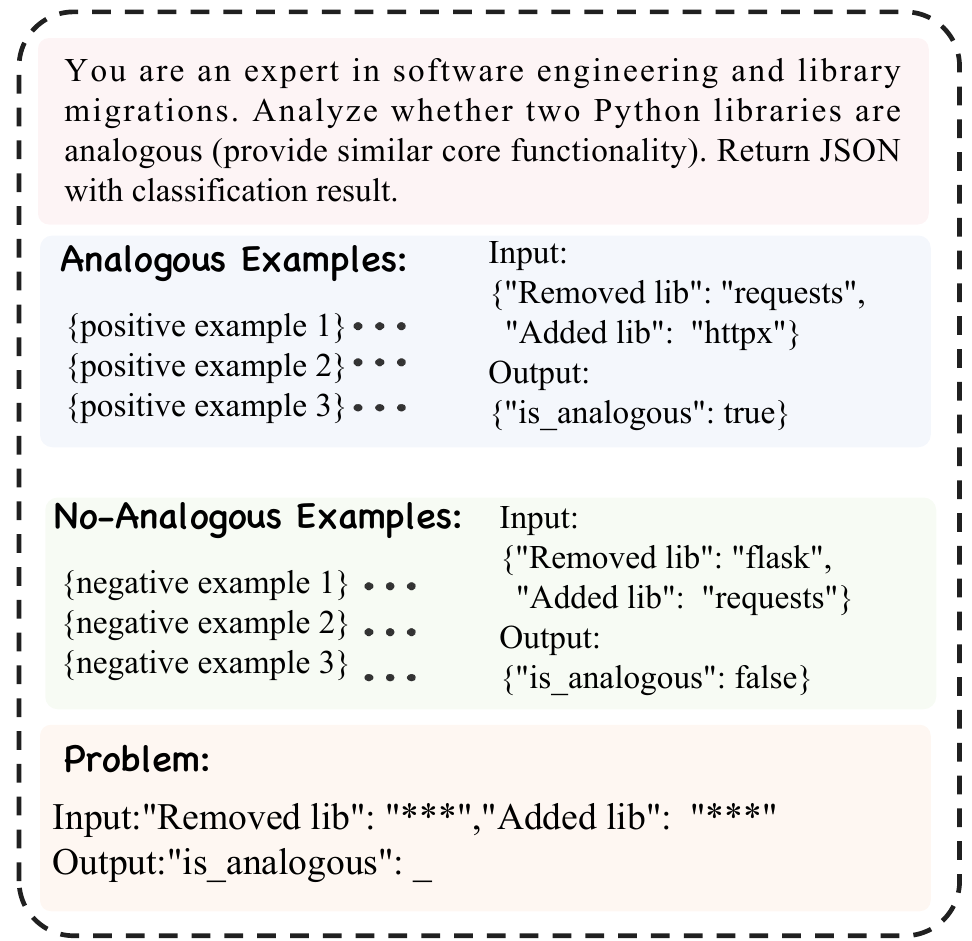}
  \caption{The prompt template used for classifying whether a library pair is analogous.}
  \label{filterprompt}
\end{figure}

To achieve this, we design an initial prompt template (Figure \ref{filterprompt}) using a few-shot strategy to guide the model effectively, and manually evaluate the effectiveness of the prompt template through an iterative process. Consequently, we identified 7,205 library pairs annotated as analogous. To assess the robustness of the LLM annotations, we randomly sample 365 examples based on a 95\% confidence level with a 5\% confidence interval \cite{boslaugh2012statistics,zhong2025ccisolver} for manual evaluation. Our manual review confirmed that the LLM annotations achieved an accuracy exceeding 98\%, demonstrating the effectiveness and robustness of the data annotations classified by the LLM (GPT-4o). Subsequently, we removed 4 library pairs where inconsistencies were identified between LLM annotations and human evaluations, resulting in 7,201 library migration rules. 

%Wang Specifically, we provide GPT-4o with examples of analogous library pairs as well as examples of irrelevant library pairs to facilitate few-shot in-context learning, enabling it to accurately label each library pair as either analogous or irrelevant. We applied random sampling to select 365 of the 7,205 library pairs annotated as analogous for manual verification at 95\% confidence level with a 5\% margin of error \cite{singh2014sampling}. The manual review confirmed that the accuracy of the annotation exceeds 98\%, resulting in the removal of 4 pairs due to inconsistencies between the human and GPT-4o annotations, thus demonstrating the effectiveness and reliability of the data annotations generated by GPT-4-0613.

\subsection{Data Generation}\label{generation}

After collecting analogous migration rules, we applied the methodology introduced in Section \ref{intent} to extract migration intents from corresponding commit messages for each library migration rule. Consequently, 2,888 high-quality migration intents were successfully generated from the commit messages. These 2,888 migration rules, along with their corresponding migration intents, were retained for further analysis. To ensure the accuracy and reliability of this approach, we randomly sampled 100 examples for manual verification. We reviewed each commit message to check the correctness of the generated intent. Our analysis confirmed that the LLM consistently produces high-quality migration intents, demonstrating its reliability and effectiveness. 

Subsequently, we applied the method described in Section \ref{intent-type} to classify migration intents into distinct intent types. To evaluate the accuracy of this classification, a human evaluation was performed on a randomly selected sample of 339 migration intents, along with their corresponding intent types. This sample was drawn from the total of 2,888 records based on a 95\% confidence level with a 5\% confidence interval \cite{boslaugh2012statistics,zhong2025ccisolver}. Human evaluation indicates that GPT-4o performs well in intent type classification, achieving an accuracy of 95\%.

\subsection{Dataset Mask and Split}\label{mask}

Following the processes of dataset collection and generation, we collected a total of 2,888 migration rules along with their corresponding migration intents and intent types. By manual observation of the dataset, we identified a potential risk of data leakage in the migration intents extracted from commit messages. Specifically, some migration intents explicitly referenced the names of target libraries or their variants, which could inadvertently disclose critical factual information when prompting LLMs. To mitigate this issue, we manually masked references to target libraries within the migration intents, replacing them with the \emph{[MASK]} tag. This ensures that the recommendations generated by LLMs are based solely on the detailed information provided, rather than relying on factual knowledge of the target libraries.

After that, the dataset was divided into training and test sets using an 8:2 ratio. To prevent any data leakage, we followed the practice in prior studies \cite{zhang2025patch}, and ensured that migration rules originating from the same repository were assigned to the same set. Additionally, all migration rules were ordered by their migration dates, with more recent rules allocated to the training set. This process yields 2,310 migration rules in the training set and 578 in the test set. Specifically, the training set serves as the primary data source for our RAG database (detailed in Section \ref{RAG}).

%% file: setup.tex
\subsection{Research Questions}

Our experiments aim to answer the following research questions (RQ):

\begin{itemize}
    \item \textbf{RQ1: Overall Performance.} How does \bench perform in library recommendation?
    \item \textbf{RQ2: Ablation Study.} How does each component contribute to the overall performance?
    \item \textbf{RQ3: Effectiveness of Different Prompt Engineering Strategies.} How do different prompt engineering strategies influence the effectiveness of our approach?
    \item \textbf{RQ4: Effectiveness across Different Intent Types.} How does \bench perform on different types of migration intents?
    \item \textbf{RQ5: Cases Study.} What are the characteristics of failed cases?
\end{itemize}

\subsection{Model Selection}
\renewcommand{\arraystretch}{1.2}
\begin{table}[ht]
\centering
\caption{Selected LLMs.}
\label{selected_llms}
\begin{tabular}{@{}lcccc@{}}
\toprule
\textbf{Model} & \textbf{Size} & \textbf{Time} & \textbf{Instruct} & \textbf{Base} \\
\midrule
GPT-4o-mini & N/A & 2024-07-18 & \ding{55} & \checkmark \\
GPT-4 & N/A & 2023-06-13 & \ding{55} & \checkmark \\
DeepSeek-V3 & 660B &2025-03-24 & \ding{55} & \checkmark \\
Qwen-Plus & N/A &2025-01-25 & \ding{55} & \checkmark \\
Llama-3.3 & 70B & 2024-12-06&\checkmark&\ding{55}\\
Claude-3.7-Sonnet & N/A & 2025-02-19&\ding{55}&\checkmark\\
Qwen-Coder-Plus & N/A & 2024-11-06 &\ding{55} &\checkmark \\
Qwen2.5-Coder &32B &2024-11-12 &\checkmark &\ding{55} \\
DeepSeek-R1 & 671B&2025-01-20 &\ding{55} &\checkmark \\
QwQ-Plus & N/A &2025-03-05 &\ding{55} &\checkmark \\
\bottomrule
\end{tabular}
\end{table}

As shown in Table \ref{selected_llms}, we evaluate the performance of ten distinct LLMs on the target library recommendation task. These models are categorized into three groups: general LLMs, code LLMs, and reasoning LLMs. Notably, we exclude GPT-4o to avoid potential data leakage, since it was used for migration intents generation. For general LLMs, we select GPT-4o-mini, GPT-4, DeepSeek-V3, Qwen-Plus, Llama-3.3, and Claude-3.7-Sonnet. Specifically, GPT-4-0613 is used as the representative version of GPT-4, while Llama-3.3-70B-Instruct serves as the implementation version for Llama-3.3. For code LLMs, we include Qwen-Coder-Plus and Qwen2.5-Coder from the Qwen family, with Qwen2.5-Coder-32B-Instruct employed as the implementation version for Qwen2.5-Coder. For reasoning LLMs, we evaluate DeepSeek-R1 and QwQ-Plus. The selection of these models is based on their outstanding performance in prior research and their availability.

\subsection{Evaluation Metrics}

Following previous studies \cite{chen2024apigen,wei2022clear}, we employ two metrics to evaluate the performance of our framework: $Precision@k$ and Mean Reciprocal Rank (MRR). 

\begin{itemize}
    \item \textbf{{$Precision@k$}} measures a model's ability to include the correct target library within the top-k results, regardless of their order. It is defined as:
    \begin{equation}
    \label{Precision@k}
    Precision@k = \frac{\sum_{i=1}^N HasTarget_k(q_i)}{N}
    \end{equation}
    where $N$ denotes the total number of queries, and $HasCorrect_k(q)$ equals 1 if the correct target library appears within the top-k results for query $q$, otherwise it returns 0.
    \item \textbf{$MRR$} is a widely adopted metric for evaluating recommendation tasks. It considers the ranks of all correct answers. It is formally defined as:
    \begin{equation}
    \label{mrr}
    MRR = \frac{\sum_{i=1}^N 1/firstpos(q_i)}{N}
    \end{equation}
    where $firstpos(q)$ returns the position of the first correct target library in the results. If the correct target library is not present in the results, the numerator of the formula is zero.
\end{itemize}

\subsection{Implementation Details}

Following Chen et al.'s study \cite{chen2025llms}, we set $k=3$ during the retrieval stage, to optimize the LLM's ability to learn from the retrieved demonstrations. Additionally, we configure the LLMs with a temperature of 0.7 and a top\_p of 0.95 to balance diversity and coherence in the generated outputs. For the evaluation metric $Precision@k$, we set $k\in \{1,3,5,10\}$ to assess the model's ability to include the correct answer within its top-k outputs.

%Wang In the retrieval stage, we set $k=3$ to maximize the LLM's ability to learn from the retrieved examples. The temperature is fixed at 0.7 and top-p at 0.95 across all experiments to ensure parameter consistency. When evaluating model performance, we computed Precision@k, with $k\in \{1,3,5,10\}$, to assess the model's ability to output the correct answer within its top-k outputs.

%% file: rq1.tex
\begin{table}[!ht]
\centering
\caption{The performance of different LLMs in our framework.}
\label{overall_performance}
\begin{tabular*}{\columnwidth}{%
      %@{\extracolsep{\fill}}           % 把剩余空间平分到各列之间
      C{2.3cm} |                                    
      C{0.8cm}  C{0.8cm}  C{0.8cm}  C{0.9cm}|C{0.7cm}                        
      %@{}                             
    }
\hline
\multirow{2}{*}{LLM} & \multicolumn{4}{c|}{Precision@k} & \multirow{2}{*}{MRR} \\
 & Top-1 & Top-3 & Top-5&Top-10&\\
\hline 
GPT-4o-mini & 0.448 & 0.611 & 0.647 & 0.673&0.532  \\ 
GPT-4 & 0.516 & 0.675 & 0.701 & 0.729&0.598  \\ 

DeepSeek-V3 &0.524 & 0.667 &0.687 &0.711&0.598  \\ 

Qwen-Plus & 0.510 & 0.638 &0.661 &0.694&0.578  \\

Llama-3.3 & 0.467 & 0.585 &0.630 &0.651&0.533  \\
Claude-3.7-Sonnet &\textbf{0.566}\cellcolor{gray!25}& \textbf{0.722}\cellcolor{gray!25} &\textbf{0.751}\cellcolor{gray!25} &\textbf{0.792}\cellcolor{gray!25}&\textbf{0.652}\cellcolor{gray!25}  \\
\hline
Qwen-Coder-Plus & 0.484 & 0.649&0.678 &0.701&0.569  \\
Qwen2.5-Coder & 0.465 & 0.633 &0.671 &0.704&0.554  \\
\hline
DeepSeek-R1 & 0.550 & 0.718 &0.730 &0.756&0.634  \\ 
QwQ-Plus & 0.521 & 0.671 &0.692 &0.735&0.598 \\ 
\hline
%Wang Average & 0.505 & 0.657 &0.685 &0.715&0.585  \\
%\hline
 \end{tabular*}
\end{table}

\begin{table*}[t]
\centering
\small
\caption{Performance of variants of \textbf{vanilla}, \textbf{w/o Ret} and \textbf{w/o Intent} compared to our framework.}
\label{ablation}
\resizebox{\textwidth}{!}{%
\begin{tabular}{l|*{10}{C{1.2cm}}L{2.5cm}}
\toprule
Setup & GPT-4o-mini & GPT-4 & DeepSeek-V3 & Qwen-Plus & Llama-3.3 & Claude-3.7-Sonnet & Qwen-Coder-Plus & Qwen2.5-Coder & DeepSeek-R1 & QwQ-Plus & Average \\
\midrule
\multicolumn{12}{c}{Precision@1} \\
\midrule
Vanilla & 0.363 & 0.405& 0.445 & 0.356 & 0.315 & 0.372 & 0.395 & 0.414 & 0.426 & 0.392 &0.388 \textcolor{red}{(↓23.17\%)} \\
w/o Ret & 0.403 & 0.484 & 0.476 & 0.464 & 0.445 & 0.514 & 0.429 & 0.436 & 0.500 & 0.460 & 0.461 \textcolor{red}{(↓8.71\%)}\\
w/o Intent & 0.396 & 0.469 & 0.388 & 0.457 & 0.360 & 0.462 & 0.460 & 0.445 & 0.424 & 0.484 &0.435 \textcolor{red}{(↓13.86\%)} \\
\textbf{LibRec} & \textbf{0.448} \cellcolor{gray!25}& \textbf{0.516}\cellcolor{gray!25} & \textbf{0.524}\cellcolor{gray!25} & \textbf{0.510}\cellcolor{gray!25} & \textbf{0.467}\cellcolor{gray!25}& \textbf{0.566}\cellcolor{gray!25} & \textbf{0.484} \cellcolor{gray!25}& \textbf{0.465}\cellcolor{gray!25} & \textbf{0.550}\cellcolor{gray!25} & \textbf{0.521} \cellcolor{gray!25}&0.505 \\
\midrule
\multicolumn{12}{c}{Precision@3} \\
\midrule
Vanilla&0.507&0.559&0.592&0.517&0.420&0.540&0.548&0.540&0.562&0.568&0.535 \textcolor{red}{(↓18.57\%)}\\
w/o Ret&0.535&0.613&0.609&0.595&0.554&0.680&0.571&0.564&0.640&0.616&0.598 \textcolor{red}{(↓8.98\%)}\\
w/o Intent&0.517&0.640&0.510&0.606&0.479&0.663&0.604&0.590&0.543&\textbf{0.680}\cellcolor{gray!25}&0.583 \textcolor{red}{(↓11.26\%)}\\
\textbf{LibRec} & \textbf{0.611}\cellcolor{gray!25} & \textbf{0.675} \cellcolor{gray!25}& \textbf{0.666}\cellcolor{gray!25} & \textbf{0.638}\cellcolor{gray!25} & \textbf{0.585}\cellcolor{gray!25} & \textbf{0.722} \cellcolor{gray!25}& \textbf{0.649}\cellcolor{gray!25} & \textbf{0.633}\cellcolor{gray!25} & \textbf{0.718}\cellcolor{gray!25} & 0.671 &0.657 \\
\midrule
\multicolumn{12}{c}{Precision@5} \\
\midrule
Vanilla&0.548&0.587&0.640&0.552&0.448&0.599&0.590&0.588&0.602&0.592&0.575 \textcolor{red}{(↓16.06\%)}\\
w/o Ret&0.564&0.640&0.640&0.623&0.592&0.716&0.614&0.602&0.675&0.640&0.631 \textcolor{red}{(↓7.88\%)}\\
w/o Intent&0.545&0.670&0.542&0.642&0.519&0.702&0.635&0.625&0.588&\textbf{0.709}\cellcolor{gray!25}&0.618 \textcolor{red}{(↓9.78\%)}\\
\textbf{LibRec}&\textbf{0.647}\cellcolor{gray!25}&\textbf{0.701}\cellcolor{gray!25}&\textbf{0.687}\cellcolor{gray!25}&\textbf{0.661}\cellcolor{gray!25}&\textbf{0.630}\cellcolor{gray!25}&\textbf{0.751}\cellcolor{gray!25}&\textbf{0.678}\cellcolor{gray!25}&\textbf{0.671}\cellcolor{gray!25}&\textbf{0.730}\cellcolor{gray!25}&0.692&0.685\\
\midrule
\multicolumn{12}{c}{Precision@10} \\
\midrule
Vanilla&0.583&0.632&0.678&0.595&0.500&0.644&0.619&0.625&0.645&0.628&0.615 \textcolor{red}{(↓13.99\%)}\\
w/o Ret&0.604&0.683&0.673&0.661&0.623&0.758&0.654&0.650&0.702&0.689&0.670 \textcolor{red}{(↓6.29\%)}\\
w/o Intent&0.573&0.711&0.568&0.670&0.562&0.744&0.670&0.668&0.625&\textbf{0.753}\cellcolor{gray!25}&0.654 \textcolor{red}{(↓8.53\%)}\\
\textbf{LibRec}&\textbf{0.673}\cellcolor{gray!25}&\textbf{0.729}\cellcolor{gray!25}&\textbf{0.711}\cellcolor{gray!25}&\textbf{0.694}\cellcolor{gray!25}&\textbf{0.651}\cellcolor{gray!25}&\textbf{0.792}\cellcolor{gray!25}&\textbf{0.701}\cellcolor{gray!25}&\textbf{0.704}\cellcolor{gray!25}&\textbf{0.756}\cellcolor{gray!25}&0.735&0.715\\
\midrule
\multicolumn{12}{c}{MRR} \\
\midrule
Vanilla &0.443&0.487&0.528&0.444&0.376&0.466&0.476&0.485&0.505&0.483&0.469 \textcolor{red}{(↓19.83\%)}\\
w/o Ret&0.473&0.555&0.549&0.537&0.506&0.603&0.508&0.510&0.577&0.545&0.536 \textcolor{red}{(↓8.38\%)}\\
w/o Intent&0.464&0.555&0.455&0.538&0.429&0.568&0.539&0.526&0.494&0.583&0.515 \textcolor{red}{(↓11.97\%)}\\
\textbf{LibRec}&\textbf{0.532}\cellcolor{gray!25}&\textbf{0.598}\cellcolor{gray!25}&\textbf{0.598}\cellcolor{gray!25}&\textbf{0.579}\cellcolor{gray!25}&\textbf{0.533}\cellcolor{gray!25}&\textbf{0.652}\cellcolor{gray!25}&\textbf{0.569}\cellcolor{gray!25}&\textbf{0.554}\cellcolor{gray!25}&\textbf{0.634}\cellcolor{gray!25}&\textbf{0.598}\cellcolor{gray!25}&0.585\\
\bottomrule
\end{tabular}%
}
\end{table*}

Table \ref{overall_performance} presents a comparative analysis of Precision@1, Precision@3, Precision@5, Precision@10, and MRR across general LLMs, code LLMs, and reasoning LLMs in the target library recommendation task. \textbf{The experimental results demonstrate that our framework can recommend target libraries effectively.} Surprisingly, among all the evaluated models, the general LLM Claude-3.7-Sonnet achieves the best performance, with the highest scores across all metrics: Precision@1 (56.6\%), Precision@3 (72.2\%), Precision@5 (75.1\%), Precision@10 (79.2\%), and MRR (65.2\%).

%Wang We evaluate our method using three categories of state-of-the-art LLMs: \textbf{(1) General LLMs.} GPT-4o-min, GPT-4, Claude-3.7-Sonnet, DeepSeek-V3, Qwen-Plus, Llama-3.3 \textbf{(2) Code LLMs.} Qwen-Coder-Plus, Qwen2.5-Coder and \textbf{(3) Reasoning LLMs.} DeepSeek-R1, QwQ-Plus. Table~\ref{tab:performance} reveals a clear hierarchy in recommendation quality across our three model categories. The results indicate that within our pipeline, state-of-the-art LLMs can efficiently prioritize and recommend the target libraries. 

Although the framework exhibits strong overall performance, there are substantial performance disparities across different categories of LLMs. Specifically, in the category of general LLMs, Claude-3.7-Sonnet outperforms the commercial models GPT-4o-mini and GPT-4 in Precision@10 by approximately 17.7\% and 8.7\%, respectively. Meanwhile, GPT-4 and DeepSeek‑V3 exhibit comparable performance in the target library recommendation task. In contrast, the open‑source general LLM Llama‑3.3 demonstrates the weakest performance in this task, which may be due to its smaller parameter size.

%Wang For example, among the general LLMs, Claude-3.7-Sonnet clearly leads. It achieves a Precision@10 of 79.2\% and an MRR of 65.2\%, which exceeds leading commercial models like GPT‑4's top‑10 precision by about 8\%. The open‑source general LLMs demonstrate a consistent but modest gap behind the leader. DeepSeek‑V3 and Qwen‑Plus deliver comparable recommendation quality, each improving steadily as more candidates are considered. Llama‑3.3 falls further behind, which may relates to its smaller parameter footprint and less frequent data refresh. 

In the category of reasoning LLMs, DeepSeek‑R1 delivers suboptimal performance across all metrics: Precision@1 (55\%), Precision@3 (71.8\%), Precision@5 (73\%), Precision@10 (75.6\%), and MRR (63.4\%). Similarly, QwQ‑Plus demonstrates relatively high performance, surpassing most other LLMs. The notable performance of DeepSeek‑R1 and QwQ‑Plus highlights the effectiveness of reasoning LLMs in addressing the target library recommendation task. 

When it comes to the category of code LLMs, the performance is observed to be inferior compared to both general and reasoning models. This indicates that code LLMs struggle to effectively recommend the correct target libraries, and lack the capability to rank relevant libraries higher in the recommendation list.

%Wang Reasoning models occupy the middle tier in our results. DeepSeek‑R1 narrows the gap to top commercial models, reaching over 75\% Precision@10 compared to Claude‑3.7‑Sonnet’s 79\%, while QwQ‑Plus follows closely behind. Both models show steadier gains beyond the first recommendation, demonstrating solid inferential strength. 

%Wang Code LLMs form the lowest tier in our comparison, lagging behind on the MRR metric compared to both general and reasoning models. Although these code‑tuned architectures deliver competitive Top‑3 precision, their lower MRR reflects a falloff in ranking quality beyond the very first suggestion. This pattern suggests that, while code‑tuned architectures can surface the most likely library early on, they lack the breadth of coverage to sustain high precision across an expanded candidate set.

%\vspace{-0.2cm}
\rqbox{\textbf{Finding 1:} Our framework can recommend target libraries effectively. Meanwhile, the general LLM Claude‑3.7‑Sonnet achieves the best performance across all metrics. At the same time, reasoning models like DeepSeek‑R1 and QwQ‑Plus demonstrate relatively suboptimal performance, highlighting the effectiveness of reasoning LLMs in addressing the target library recommendation task.}
%\vspace{-0.2cm}

%% file: rq2.tex
\subsubsection{Setup} 

\begin{figure*}[!t]
  \centering
\includegraphics[width=\textwidth,keepaspectratio]{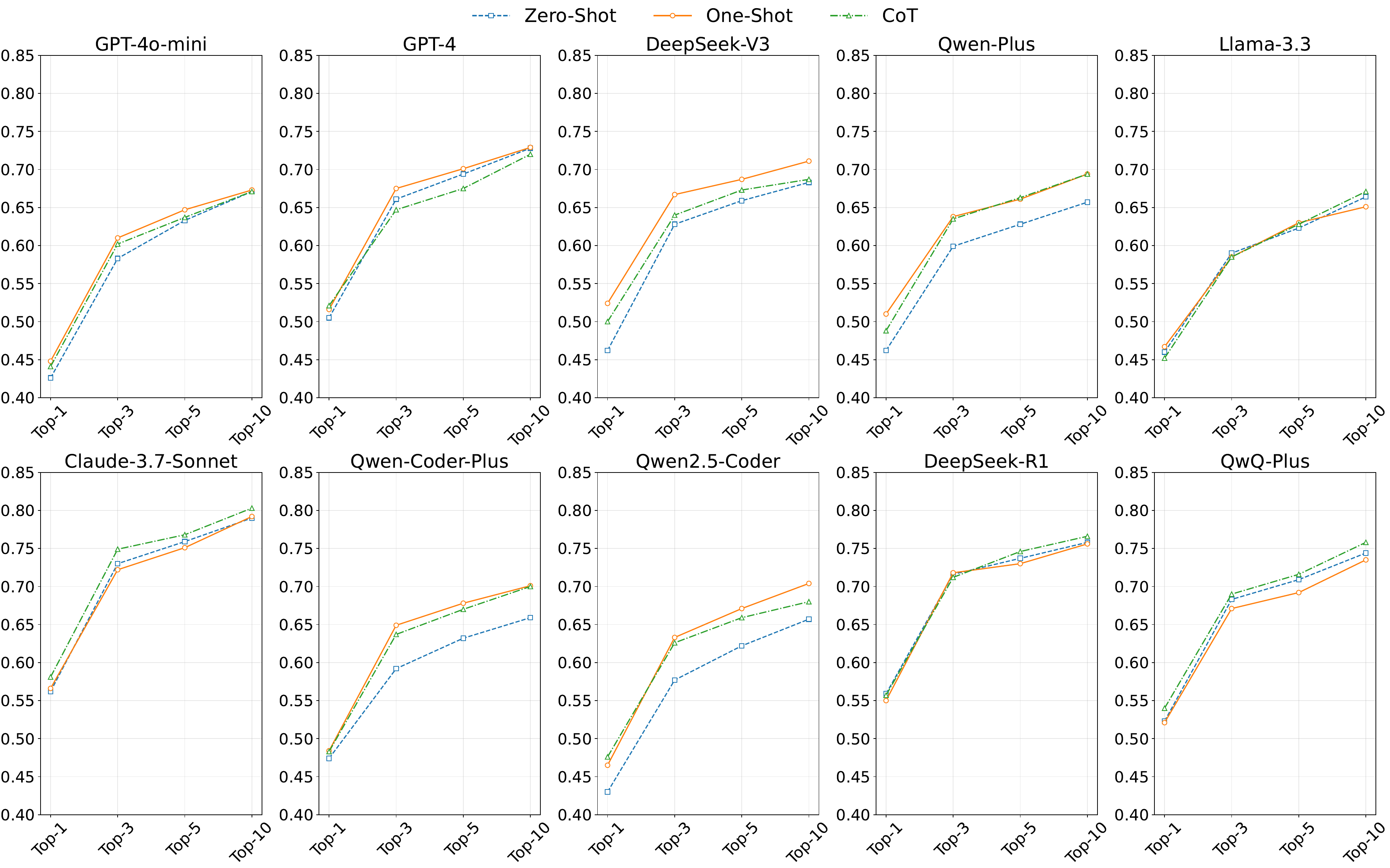}
  \caption{Precision@K Trends of Various LLMs Across Different Prompt Strategies.}
  \label{fig_rq3}
\end{figure*}

%Wang To answer this question, we conduct a series of ablation experiments to evaluate the impact of various components within the pipeline design. To ensure the fairness of comparisons, we maintained consistency in the implementation settings. Table \ref{tab:ablation} presents the evaluation results, where each row for a given LLM corresponds to one ablation variant. The symbol \emph{w/o Ret} denotes the removal of the retrieval module, \emph{w/o Intent} denotes the exclusion of migration intents from both input and ranking, and w/Ret Intent indicates the full pipeline of our method. In addition, we include a vanilla baseline for each model, which only relies on the ability of LLM without any retrieval or intent guidance, serving to illustrate the gains contributed by each pipeline component.

In this section, we conduct ablation studies to explore the necessity of different components for the performance of our framework. To ensure fairness, all implementation settings were kept consistent across experiments. As shown in Table \ref{ablation}, we introduce several ablation variants: \textbf{w/o Ret} denotes the removal of the retrieval-augmented component, \textbf{w/o Intent} represents the exclusion of migration intents from the prompt, i.e., remove the context enhancement brought by migration intents, and \textbf{\bench} refers to our complete recommendation framework. Additionally, we provide a \textbf{vanilla} baseline for each model, which only relies on the ability of LLMs without incorporating retrieval or context enhancement. These ablation studies allow us to systematically analyze the contribution of each component to the overall performance of our proposed framework.

\subsubsection{Results}

As shown in Table \ref{ablation}, under \textbf{vanilla} conditions, the average scores for Precision@1, Precision@3, Precision@5, Precision@10, and MRR are 38.8\%, 53.5\%, 57.5\%, 61.5\%, and 46.9\%, respectively. The relatively low scores highlight the inherent difficulty models face in recommending target libraries based solely on their native capabilities. Subsequently, under \textbf{w/o Ret} condition, the average scores are 46.1\%, 59.8\%, 63.1\%, 67\%, and 53.6\%, reflecting declines of 8.71\%, 8.98\%, 7.88\%, 6.29\%, and 8.38\%, respectively, compared to the complete framework. Similarly, under \textbf{w/o Intent} condition, the average scores are 43.5\%, 58.3\%, 61.8\%, 65.4\%, and 51.5\%, corresponding to declines of 13.86\%, 11.26\%, 9.78\%, 8.53\%, and 11.97\%, respectively, compared to the complete framework. \textbf{These results give a hint that all key components are essential for achieving the best effectiveness.} 

Moreover, we observe that, on average, when generating only the top-1 result, migration intents are more critical than retrieval augmentation. However, as the number of output results increases, the significance of retrieval augmentation surpasses that of migration intents. Interestingly, different LLMs behave differently. For instance, for most LLMs, including GPT-4o-mini, GPT-4, Llama-3.3, Qwen-Coder-Plus, Qwen2.5-Coder, DeepSeek-R1, and QwQ-Plus, the retrieval augmentation component demonstrates greater importance across most evaluation metrics compared to migration intents. However, several LLMs of DeepSeek-V3, Qwen-Plus, and Claude-3.7-Sonnet exhibit the opposite trend. This phenomenon may be attributed to the fact that the LLMs of DeepSeek-V3, Qwen-Plus, and Claude-3.7-Sonnet may possess stronger capabilities in contextual understanding, enabling them to extract more meaningful information from migration intents, thereby reducing their reliance on retrieval augmentation. 

%WANG Without retrieval, Top‑1 precision falls by about 4\% on average. The decline at MRR approaches 5\% across models. This pattern shows that retrieval is crucial for identifying a broad set of relevant migration rules. Without intent guidance, the model still benefits from relevant retrieved candidates and maintains comparable performance at broader cutoffs. However, we observe consistent declines in early-stage metrics such as Precision@1 and MRR, suggesting that intents play a critical role in helping the model rank the most suitable libraries at the top. 

%WANG Comparing all variants against the full pipeline underscores the importance of each module. Retrieval delivers the largest gains at higher $k$ values and ensures a comprehensive candidate pool. Intents sharpen early ranking and improve overall score consistency. Together, these components make our pipeline significantly more effective than either a standalone design or a vanilla prompt approach.

\rqbox{\textbf{Finding 2:} Both the retrieval-augmented component and migration intents within our framework are essential for achieving the best effectiveness. On average, retrieval augmentation demonstrates greater significance compared to migration intents; however, the relative importance of these components varies across different LLMs.}

%% file: rq3.tex
%Wang As shown in Figure \ref{fig_rq3}, the results demonstrate performance variations across three prompt paradigms in library recommendation tasks. One‑Shot delivers the most consistent results across six models. It achieves the highest scores in GPT‑4‑0613 and GPT‑4o‑mini. It also leads in DeepSeek‑V3, Qwen‑Plus, Qwen‑Coder‑Plus, and Qwen2.5‑Coder. The advantage is most evident at early cutoffs but remains robust at Top‑10. In contrast, Zero‑Shot yields more variable outcomes. In contrast, Zero‑Shot lags by approximately 4\% behind both One‑Shot and COT on Qwen‑Plus and Qwen2.5‑Coder. It nonetheless slightly outperforms those paradigms on Llama‑3.3 and DeepSeek‑R1. COT excels on Claude‑3.7‑Sonnet and QwQ‑Plus where it surpasses both other paradigms. It also ranks among the top two paradigms across all models. These observations indicate that One‑Shot is the most reliable prompt design and that COT can provide additional gains for certain high‑capacity models.

\begin{figure*}[!ht]
  \centering
\includegraphics[width=\textwidth,keepaspectratio]{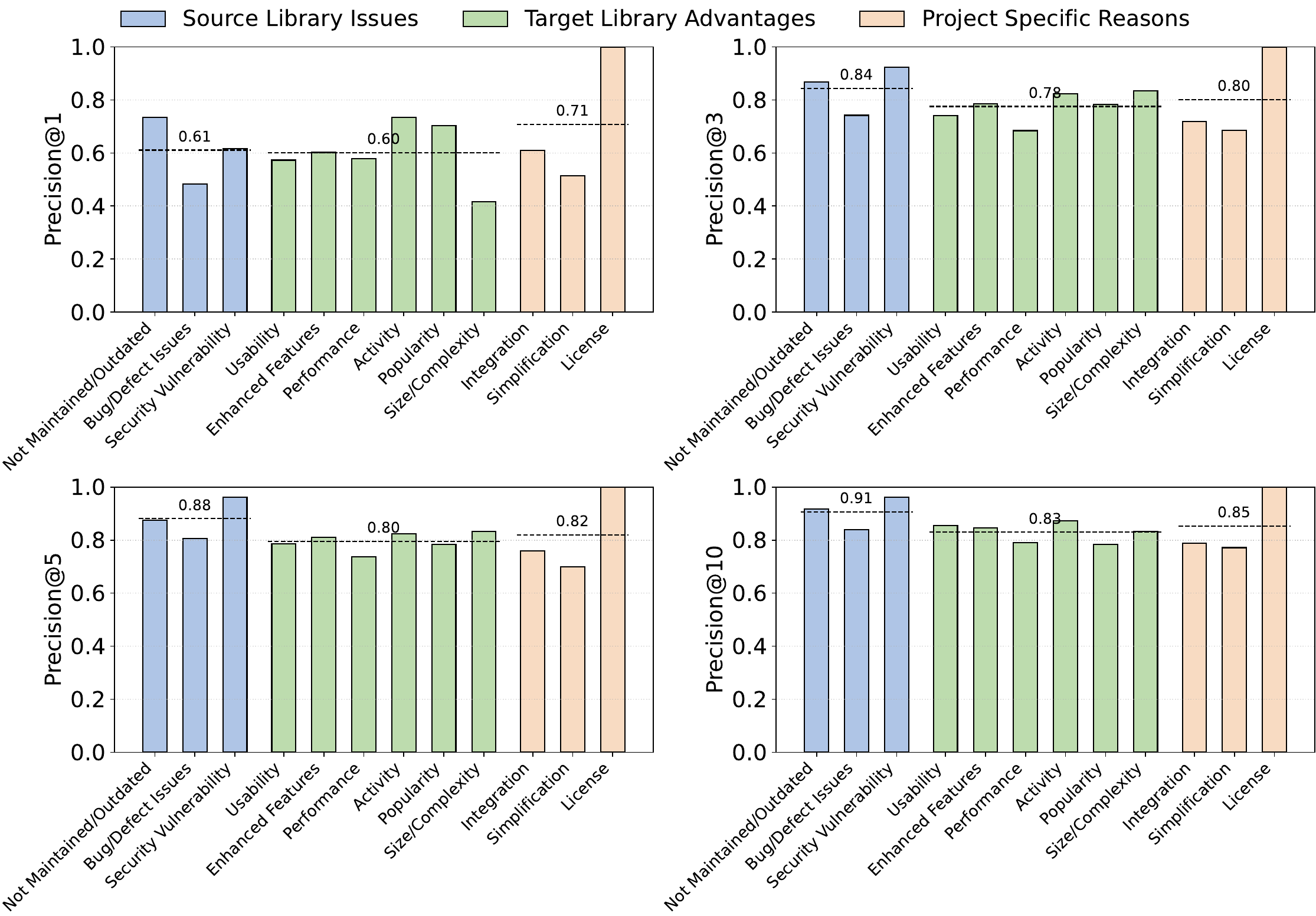}
  \caption{The effectiveness of our framework (take the best-performing LLM of Claude‑3.7‑Sonnet as an example) across various intent types. The horizontal axis represents all sub-categories of intent types, and different colors indicate distinct categories. The values on the dotted line denote the average score of our framework in each category.}
  \label{tab:rq4}
\end{figure*}

As shown in Figure \ref{fig_rq3}, the One-Shot prompt strategy demonstrates superior performance across 6 models, including 4 general LLMs of GPT‑4o‑mini, GPT-4, DeepSeek-V3, Qwen‑Plus, and 2 code LLMs of Qwen‑Coder‑Plus and Qwen2.5‑Coder. Among them, the One-Shot strategy achieves the highest Precision@10 scores in GPT-4, DeepSeek-V3, reaching 72.9\% and 71.1\%, respectively. \textbf{These results confirm that the One‑Shot strategy is the most reliable prompt strategy for our task}, as it provides a concrete example to reduce model burden, thereby helping LLMs to generate target libraries more effectively.

In contrast, the CoT prompt strategy exhibits superior performance in only 3 models: 1 general LLM of Claude‑3.7‑Sonnet, and 2 reasoning LLMs of DeepSeek‑R1 and QwQ‑Plus. Among these, the CoT strategy achieves its highest Precision@10 scores of 80.3\% with Claude‑3.7‑Sonnet. This phenomenon can be attributed to the fact that reasoning LLMs, such as DeepSeek-R1 and QwQ-Plus, are typically optimized for logical reasoning tasks during their training process, endowing them with stronger logical analysis and step-by-step reasoning capabilities. Consequently, the CoT strategy can better unlock the potential of these models, enabling superior performance in multi-step problem-solving tasks. Furthermore, the design characteristics of Claude-3.7-Sonnet allow it to maintain logical coherence more effectively when addressing complex reasoning tasks, which may contribute to its excellent performance under the CoT strategy.

\rqbox{\textbf{Finding 3:} The One-shot prompt strategy emerges as the most robust and reliable paradigm across six models, consistently demonstrating superior performance. In contrast, the CoT prompt strategy excels primarily in reasoning LLMs and high-capacity models like Claude-3.7-Sonnet.}

%% file: rq4.tex
In this RQ, we analyzed the target library recommendation performance of our framework across different intent types, using the Claude‑3.7‑Sonnet model as an example, as illustrated in Figure \ref{tab:rq4}. Figure \ref{tab:rq4} reports the Precision@1, Precision@3, Precision@5, and Precision@10 scores of our framework for each intent category and its subcategories. Among these, the category \emph{Source Library Issues} achieves the highest average scores of 84\%, 88\%, and 91\% for precision@3, precision@5, and precision@10, respectively, followed by \emph{Project Specific Reasons}. However, the category \emph{Target Library Advantages} exhibits the lowest average scores. The superior performance of \emph{Source Library Issues} may be attributed to its clear and concise description when a library is outdated or vulnerable. Conversely, the relatively lower performance of \emph{Target Library Issues} could be due to the large number of subcategories, which may require more detailed contextual information to achieve accurate target library ranking recommendations.

Within the \emph{Source Library Issues} category, our framework performs better in the subcategory \emph{Security Vulnerability}, achieving Precision@3, precision@5, and precision@10 scores of 92\%, 96\%, and 96\%, respectively. However, the performance is the weakest in the subcategory \emph{Bug/Defect Issues}, with scores of 74\%, 81\%, and 84\%, reflecting declines of 18\%, 15\%, 12\%, respectively, compared to the best-performing subcategory.

Within the \emph{Project Specific Reasons} category, our framework demonstrates superior performance in the subcategory \emph{License}, achieving near-perfect accuracy across all metrics. This strong performance can be attributed to the binary nature of licensing requirements, which impose clear constraints that our search and ranking pipeline can effectively leverage to recommend the correct target libraries with high confidence. As for the \emph{Target Library Advantages} category, our framework performs better in the subcategories of \emph{Size/Complexity} and \emph{Activity}, both achieving Precision@3 and Precision@5 scores of 82\% and 83\%, respectively. However, the performance is notably weaker in the subcategory \emph{Performance}, with Precision@3, precision@5, and precision@10 scores of 68\%, 74\%, and 79\%, respectively.

\rqbox{\textbf{Finding 4:} Our framework demonstrates superior performance in the \emph{Source Library Issues} category, followed by \emph{Project Specific Reasons}, while exhibiting the lowest performance in the \emph{Target Library Advantages} category. Notably, within the \emph{Source Library Issues} category, the framework excels in the sub-category \emph{Security Vulnerability}. Similarly, in the \emph{Project Specific Reasons} category, the framework attains near-perfect accuracy across all metrics in the sub-category \emph{License}. However, its performance is comparatively weaker in the subcategory \emph{Performance} within the \emph{Target Library Advantages} category.}

%Wang Recommendations driven by clear \emph{source‑library issues} consistently achieve the strongest results, reflecting the effectiveness of our retrieval and ranking when clear problems (such as deprecation or vulnerabilities) are present. \emph{Advantages of target libraries} and \emph{project‑specific motivations} remain more challenging and improve gradually as more candidates are considered.}

%% file: rq5.tex
%Wang In RQ5, we examine the characteristics of cases that our pipeline failed to rank within the top 10. Figure \ref{rq5timeline} shows that failures peak around 2016, reaching nearly forty cases at that point. This peak coincides with a wave of major version upgrades across many libraries, which introduced breaking changes and left migration examples sparse or inconsistent. 

\begin{figure}[!htbp]
  \centering
\includegraphics[width=1.0\columnwidth,keepaspectratio]{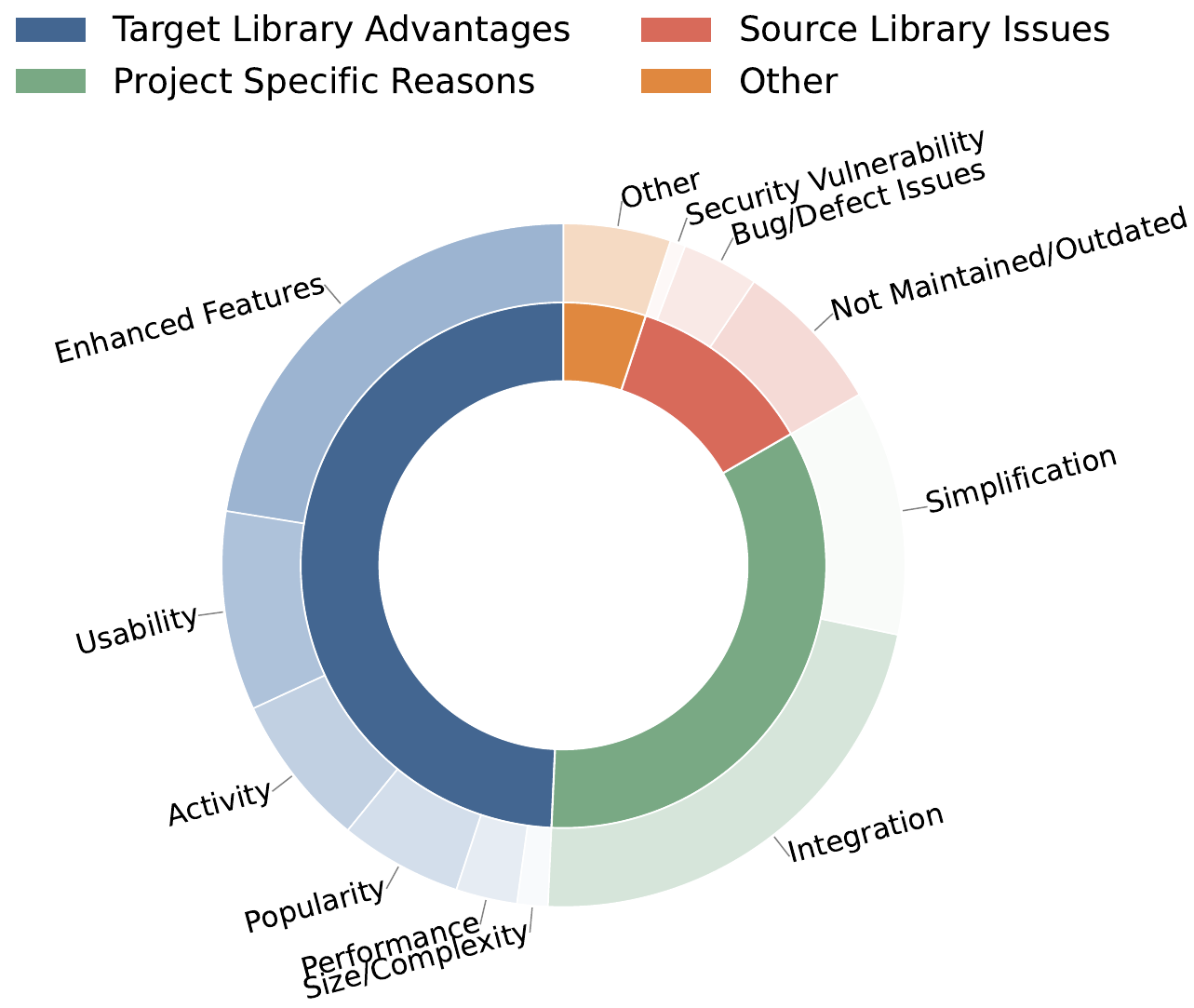}
  \caption{The distribution of intent types among failed cases.}
  \label{rq5fenlei}
\end{figure}

\begin{figure}[!htbp]
  \centering
\includegraphics[width=1.0\columnwidth,keepaspectratio]{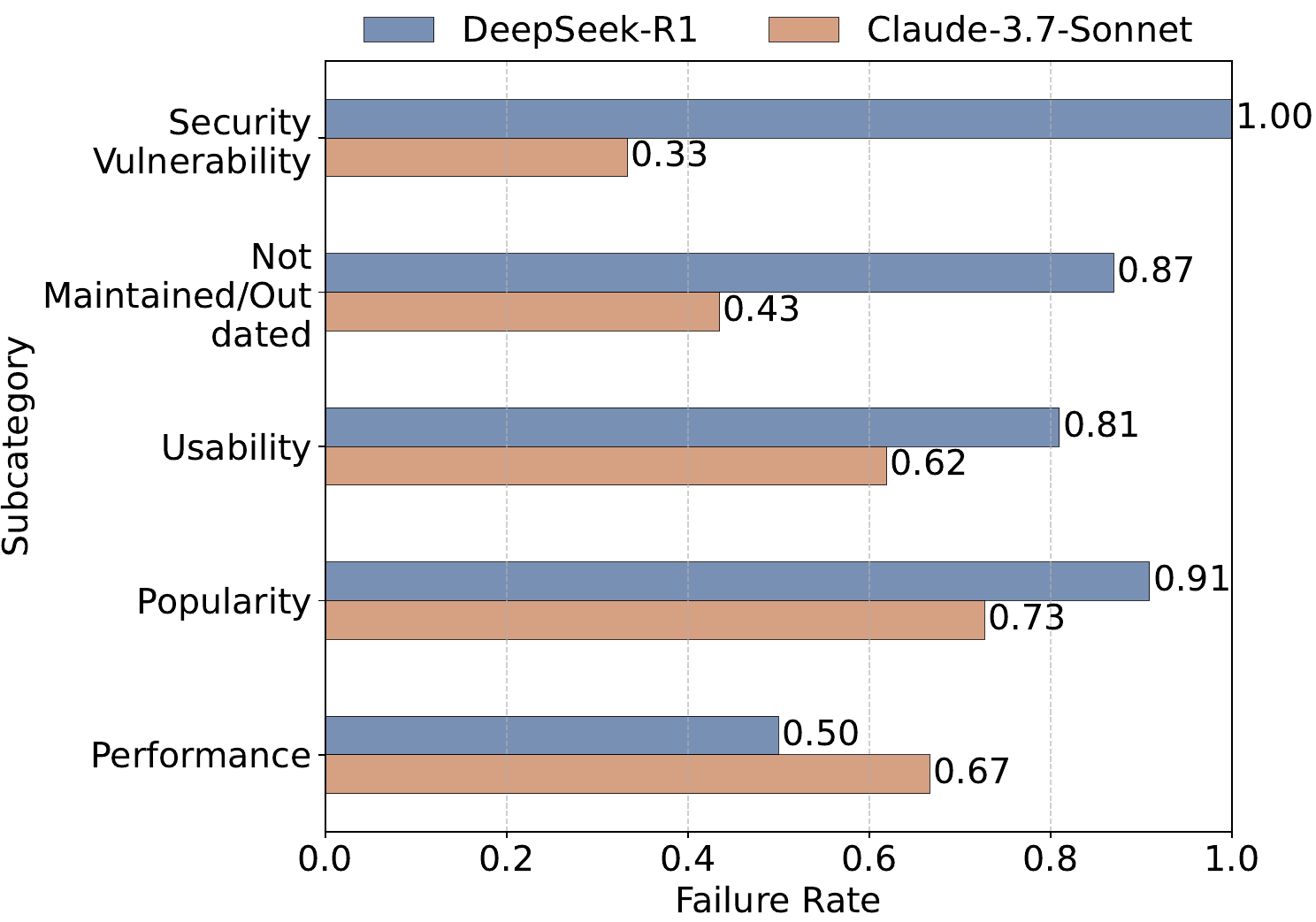}
  \caption{Comparison of failure rates for Claude-3.7-Sonnet and DeepSeek-R1 when recommending target libraries across different subcategories (top 5 subcategories with the most significant performance disparities).}
  \label{rq5difference}
\end{figure}

In this RQ, we analyze the characteristics of failed cases. Figure \ref{rq5fenlei} illustrates the distribution of intent types among these failures. Consistent with the findings of RQ4, most of the recommendation failures fall into the \emph{Target Library Advantage} category, while the \emph{Source Library Issues} category exhibits the fewest failures. Notably, within the \emph{Target Library Advantage} category, the \emph{Enhanced Features} sub-category represents the largest proportion, comprising 46\% of the total failures in this group. This may indicate that recommending target libraries based on subtle feature improvements remains a significant challenge for LLMs. For instance, in the project \emph{movies-python-bolt}, under the migration intent \emph{Update Flask demo app to use bolt driver}, the developer replaces the source library \texttt{neo4jrestclient} with the target library \texttt{neo4j-driver}. 

In contrast, within the \emph{Source Library Issues} category, the \emph{Security Vulnerability} sub-category accounts for the smallest proportion, representing only 6\% of the total failures in this group. For example, in the project \emph{steam} with the migration intent \emph{replace cryptography with pycryptodomex}, the developer replaces the source library \texttt{cryptography} with the target library \texttt{pycryptodomex}. Meanwhile, in the \emph{Project Specific Reasons} category, the \emph{Integration} sub-category constitutes the largest proportion, accounting for 66\% of the total failures in this group. For instance, in the project \emph{pax}, under the migration intent \emph{'Better integrate Pyxie', 'Better utilize Configuration system'}, the developer replaces the source library \texttt{configglue} with the target library \texttt{confiture}.

%Wang Figure \ref{rq5fenlei} breaks down the failure distribution by subcategory. Within \emph{Target Library Advantages}, the \emph{Enhanced Features} subcategory accounts for over thirty percent of that group’s failures. This likely reflects the challenge of inferring nuanced feature improvements from a limited migration context. By comparison, \emph{Size/Complexity} contributes only a small fraction, indicating that volumetric metrics are easier to capture. \emph{Security Vulnerability} also represents a minor share of Source Library Issues failures, perhaps because vulnerability descriptions carry explicit terminology that our pipeline handles well.

Furthermore, our findings in RQ1 reveal that the overall performance of Claude-3.7-Sonnet and DeepSeek-R1 is comparable. To gain deeper insights, we further analyze their failure rates when recommending target libraries across different sub-categories. Specifically, we calculated the failure rates for both models within each subcategory, and identified the five subcategories with the most significant performance disparities, as depicted in Figure \ref{rq5difference}. Our analysis shows that in the \emph{Security Vulnerability} subcategory, DeepSeek-R1 exhibits a 100\% failure rate, whereas Claude-3.7-Sonnet demonstrates a lower failure rate of 33\%. This indicates that Claude-3.7-Sonnet is more adept at recommending target libraries for migration intents involving \emph{Security Vulnerability}. For the \emph{Not Maintained/Outdated}, \emph{Usability}, and \emph{Popularity} subcategories, DeepSeek-R1's failure rates exceed 81\%, surpassing those of Claude-3.7-Sonnet by 44\%, 19\%, and 18\%, respectively. However, it is worth noting that Claude-3.7-Sonnet also exhibits a relatively high failure rate in the \emph{Popularity} subcategory. 

Interestingly, in the \emph{Performance} subcategory, DeepSeek-R1 demonstrates a relatively low failure rate of 50\%, whereas Claude-3.7-Sonnet shows a higher failure rate in this category. This suggests that DeepSeek-R1 is more effective at recommending target libraries with migration intent of \emph{Performance}. These findings indicate that although the two models achieve similar overall performance, their capabilities in recommending target libraries vary across different migration intents, highlighting the strengths and weaknesses of each model in specific contexts.

\rqbox{\textbf{Finding 5:} Most of the recommendation failures fall into the \emph{Target Library Advantage} category, with \emph{Enhanced Features} sub-category representing the largest proportion of failures within this group. Conversely, the \emph{Source Library Issues} category accounts for the fewest failures, with the \emph{Security Vulnerability} sub-category making up the smallest proportion within this category. In most subcategories of intent types, DeepSeek-R1 demonstrates a higher failure rate compared to Claude-3.7-Sonnet, while it shows a lower failure rate in the \emph{Performance} subcategory.}

%Wang Failure rates peaked during periods of major version overhauls, and feature‑enhancement migrations account for the majority of missed top‑10 suggestions, indicating areas where richer contextual understanding is needed. Also, although DeepSeek-R1 and Claude-3.7-Sonnet exhibit similar overall performance, they demonstrate notable differences in handling specific types of recommendation failures. This indicates that each model may be better suited to certain scenarios or challenges, potentially due to differences in their training data.}

%% file: implication.tex
In this section, we discuss the implications of our study and propose recommendations for researchers, software developers, and users.

%Wang \textbf{Reliable Foundations for Library Recommendation Systems.} 
%Wang We provide a dataset containing 2,888 real-world library migration records, each of which has been manually verified and enriched with detailed migration intents and categorized motivations. These additional annotations are missing from existing datasets and allow for a more fine-grained, intent-aware analysis of migration behavior. Unlike previous datasets that only provide library pairs, our dataset includes rich contextual information that reflects developers’ actual migration goals. This makes it particularly suitable for research on intent-driven library recommendation. Furthermore, our findings in RQ1 demonstrate that the proposed method achieves strong accuracy across different types of large language models. This suggests that our pipeline is robust and can serve as a reliable foundation for future work in library mapping and recommendation. Researchers can build upon our method, adapting it to other programming languages and ecosystems to broaden its applicability.

\textbf{Our findings in RQ1} suggest that the superior performance of Claude-3.7-Sonnet across all metrics establishes it as a benchmark model within our framework for library recommendation tasks. \emph{Researchers} can leverage this model as a reference point for evaluating new models or approaches in this domain. Furthermore, the relatively high performance of reasoning LLMs like DeepSeek‑R1 and QwQ‑Plus highlights the importance of reasoning capabilities in addressing library recommendation tasks. This finding highlights an opportunity for researchers to investigate how reasoning architectures or training strategies can further enhance performance in recommendation tasks. In contrast, the inferior performance of code LLMs reveals an avenue for future research aimed at improving their ability to rank and recommend libraries effectively, thereby enabling them to recommend suitable libraries in code generation tasks.

For \emph{software developers}, prioritizing the adoption of Claude-3.7-Sonnet is strongly recommended, as it demonstrates the best overall performance. Nonetheless, reasoning LLMs like DeepSeek‑R1 and QwQ‑Plus may be a viable alternative due to their comparable performance in the target library recommendation task. Conversely, the inferior performance of code LLMs indicates that they may not be suitable for tasks that require ranking or recommending target libraries. Developers should avoid relying solely on code LLMs for such tasks and instead consider integrating them with general or reasoning models to achieve superior performance. 

For \emph{users}, it is essential to be aware of the distinct strengths and limitations of general, reasoning, and code LLMs, especially that the Claude-3.7-Sonnet currently offers the highest performance for library recommendation tasks. This understanding will enable users to make more informed decisions when selecting libraries or tools for their needs.

\textbf{Our findings in RQ2} highlight the significance of both the retrieval augmentation and migration intents components in our framework. To achieve optimal effectiveness in library recommendation tasks, researchers are encouraged to integrate both components. Besides, \emph{both researchers and software developers} should be aware of the specific characteristics of the LLMs they employ. For instance, models like DeepSeek-V3, Qwen-Plus, and Claude-3.7-Sonnet may benefit more from enhanced migration intents, whereas others might rely more heavily on retrieval augmentation. In this regard, they could explore combining LLMs with diverse capabilities to further enhance the overall performance of library recommendation tasks.

\textbf{Our findings in RQ3} indicate that the One-Shot prompt strategy demonstrates superior effectiveness across a wider range of models for library recommendation tasks. \emph{Researchers} are encouraged to prioritize this strategy when developing or evaluating new models to enhance overall performance. Conversely, the superior performance of the COT strategy with reasoning LLMs like DeepSeek-R1 and QwQ-Plus highlights the importance of tailoring prompt strategies to align with the strengths of specific models. Researchers could focus on optimizing reasoning LLMs to fully exploit the potential of the CoT strategy.

Given that different prompt strategies play distinct roles in different types of LLMs, it is recommended that \emph{developers and users} select prompt strategies based on the characteristics of the models they employ. For general and code LLMs, the One-Shot strategy may yield better results, while the CoT strategy could be more beneficial for reasoning LLMs.

\textbf{Our findings in RQ4} reveal that the performance of our framework varies across different intent categories and sub-categories. \emph{Researchers and developers} should focus on optimizing recommendation frameworks for intent types that currently exhibit weaker performance. For instance, the framework demonstrates weaker performance in the \emph{Target Library Advantages} category and its sub-category \emph{Performance}. To address these limitations, researchers and developers could incorporate more detailed contextual information into prompts, or augment the models with additional data sources to better address these issues. 

Furthermore, to improve the accuracy of target library recommendations, \emph{users} should provide clear and concise contextual information, especially when dealing with intents such as \emph{Target Library Advantages}, where detailed inputs are critical for achieving better performance. Additionally, users should critically evaluate the recommendation results for weaker-performing sub-categories, such as \emph{Performance} or \emph{Bug/Defect Issues}, and consider supplementing the system's outputs with manual analysis or verification.

\textbf{Our findings in RQ5} highlight that the high proportion of failures under the \emph{Enhanced Features} subcategory within the \emph{Target Library Advantage} category underscores the challenge of recommending libraries based on subtle feature improvements. \emph{Researchers} should focus on developing advanced methods to better capture nuanced differences between libraries, potentially by incorporating advanced feature-comparison techniques or leveraging fine-grained contextual embeddings, to enhance the recommendation performance. Besides, our findings unveil significant differences in failure rates between Claude-3.7-Sonnet and DeepSeek-R1 across various subcategories. In this regard, researchers should investigate the underlying mechanisms that contribute to these disparities. For example, Claude-3.7-Sonnet's superior performance in the \emph{Security Vulnerability} subcategory suggests that it is better equipped to handle well-defined and critical issues. Researchers could analyze and explore ways to replicate these strengths in other models to improve their performance.

For \emph{developers}, they could explore the possibility of creating hybrid recommendation frameworks that leverage the strengths of both Claude-3.7-Sonnet and DeepSeek-R1. For instance, the system could dynamically switch between models according to the migration intent, using Claude-3.7-Sonnet for \emph{Security Vulnerability} and DeepSeek-R1 for \emph{Performance}.

Moreover, to enhance recommendation accuracy, \emph{users} should provide as much detailed contextual information as possible, particularly for intents such as \emph{Enhanced Features}, where subtle differences between libraries are critical. Clear descriptions of the desired features or improvements can help the recommendation framework better understand user needs and generate more accurate and reliable results.

%% file: threat.tex
%external % 数据集规模；  数据集语言Python的可扩展性；   LLM的使用 提出framework  是微调LLM的下限   另外3类LLM均有涉及，确保LLM能力调试的广泛性
%internal intent type的确定

\textbf{Internal Validity.} Our primary threat to internal validity relates to the fact that we only extract candidate migration commits from the \emph{requirement.txt} dependency file. Given that Python projects may use other types of dependency files, such as setup.py, and that dependency files are often manually named and maintained, there exists a risk of omissions in our collection of candidate migration commits. However, prior research \cite{han2020empirical}, indicates that the majority of configuration information in Python projects is typically stored in the \emph{requirements.txt} file. Moreover, prior studies \cite{han2020empirical} have also only relied on \emph{requirement.txt} to extract library dependencies. In this regard, our threat to internal validity can be mitigated to a certain degree. 

%Wang Threats to internal validity arise from experimenter bias and errors. Our approach for collecting historical migration records has the following limitations. (1) We removed forks to avoid duplicate commit analysis, but forks can contain commits not present in the original repository. (2) We only detect library additions and deletions via updates to files such as requirements.txt. Because these files are manually named and maintained, migrations that do not appear there may be missed. (3) We identify migration pairs by looking for additions and deletions within the same commit, yet real migrations sometimes span multiple commits. (4) To reduce noise, we discarded dependency‑change events involving massive or highly unbalanced edits. This may have removed some valuable migration examples. Despite these limitations, they do not undermine the validity of our findings. Our primary goal is to prompt the LLM to learn recurring historical migration patterns, not to conduct an exhaustive capture of every migration event on GitHub.

%\textbf{Construct Validity.}

\noindent \textbf{External Validity.} Threats to external validity pertain to the generalizability of our results. The first threat concerns the representativeness of our constructed benchmark with respect to the entire Python library ecosystem. Although it is possible that our benchmark does not encompass all library migrations within the ecosystem, we conducted a large-scale study involving 62,986 Python repositories and collected a total of 2,888 records for our benchmark. We believe it is a good representation of the Python ecosystem. 

The second threat relates to the applicability of our results beyond Python. However, Python has emerged as the most popular programming language on GitHub since 2024, and is widely adopted in AI-related domains. As a result, our study can bring valuable insights for the AI field and can serve as a reference for similar research conducted in other programming languages. 

The final threat pertains to the use of LLMs. To strengthen the reliability of our results, we employed three categories of LLMs, including general LLMs, reasoning LLMs, and code LLMs, encompassing a total of ten models. This comprehensive approach ensures a more robust evaluation of their capabilities in our task.

%Wang Threats to external validity concern the generalizability of our results. Our study focuses solely on Python library migrations, limiting its applicability to ecosystems such as npm or Maven. By drawing data from popular open‑source projects, we may overlook the migration behaviors found in niche or private repositories. Existing heuristic methods rely on rules tailored to specific datasets and do not generalize across different databases, and since no prior work has used large language models for library‑migration recommendations, we have not compared our approach directly with those techniques.

%% file: conclusion.tex
%Wang This paper presents a study on recommending candidate target libraries while migrating python libraries with intents. We constructed a specialized dataset of 2,888 library pairs drawn from real-world historical migration records, with each entry annotated by migration intents and intent categories extracted from its commit message. We employ a retrieval-augmented generation approach using large language models to recommend target libraries, and then evaluated the performance of our approach in library migration task (RQ1), the ablation study (RQ2), the impact of prompt paradigms (RQ3), the performance across intent types (RQ4), and the failed cases study (RQ5). Our results demonstrate that combining retrieval and intent guidance significantly enhances LLM‑based migration recommendations, yielding robust top‑k accuracy across diverse scenarios. Consequently, we generated significant findings, and provided practical implications for developers, users, and researchers. Future work will extend our pipeline to support one‑to‑many and many‑to‑many migrations, integrate automated code refactoring to enact suggested migrations. We also plan to evaluate cross‑ecosystem generality (e.g., PyPI, Maven) and to incorporate user‑in‑the‑loop feedback to further refine recommendation quality and adoption in real‑world development environments.

In this paper, we propose \bench, a framework that integrates the capabilities of LLMs with RAG techniques and in-context learning of migration intents to automate the recommendation of target libraries. Additionally, we present \data, a novel benchmark designed to evaluate the performance of target library recommendation tasks. Leveraging \data, we evaluated the effectiveness of ten popular LLMs within our framework (RQ1), conducted an ablation study to assess the contributions of key components (RQ2), analyzed the impact of various prompt strategies on the framework's performance (RQ3), evaluated its effectiveness across different intent types (RQ4), and performed detailed failure case analyses (RQ5). Our experimental findings demonstrate that \bench achieves overall optimal performance in recommending target libraries, with both retrieval-augmentation and migration intent components being critical for optimal effectiveness. The one-shot prompt strategy emerges as the most robust and reliable paradigm across six LLMs, consistently delivering superior performance. Furthermore, the framework exhibits superior performance for intent types related to issues with source libraries. 

In future work, we plan to incorporate more LLMs in our framework and design another framework to generate migration code automatically based on the recommended libraries of our framework in this paper.